\documentstyle[aps,epsf]{revtex}
\tighten
\newcommand{\xbj}{x_{\scriptscriptstyle B}}

\newcommand{\st}{{\scriptscriptstyle T}}
\newcommand{\bkt}{\bbox{k}_{\scriptscriptstyle T}}
\newcommand{\bk}{\bbox{k}}
\newcommand{\bx}{\bbox{x}}
\newcommand{\bxt}{\bbox{x}_{\scriptscriptstyle T}}
\newcommand{\bl}{\bbox{l}}

\newcommand{\bS}{\bbox{S}}

\newcommand{\baa}{\bbox{a}}
\newcommand{\bbb}{\bbox{b}}
\newcommand{\bqt}{\bbox{q}_{\scriptscriptstyle T}}
\newcommand{\bp}{\bbox{p}}
\newcommand{\bnull}{\bbox{0}}
\newcommand{\bpt}{\bbox{p}_{\scriptscriptstyle T}}
\newcommand{\bP}{\bbox{P}}

\newcommand{\bhh}{\hat{\bbox{h}}}

\newcommand{\htF}{\tilde D} 
\newcommand{\htG}{\tilde G} 
\newcommand{\htH}{\tilde H}
\newcommand{\htE}{\tilde E}
\newcommand{\hF}{D}
\newcommand{\hG}{G}
\newcommand{\hH}{H}
\newcommand{\hE}{E}
\newcommand{\ba}{\begin{eqnarray}}
\newcommand{\ea}{\end{eqnarray}}
\newcommand{\be}{\begin{equation}}
\newcommand{\ee}{\end{equation}}

\begin{document}
 
\preprint{\parbox[b]{3.3 cm} {NIKHEF 95-053\\hep-ph/9510301}}
 
\draft
 
\title{The complete tree-level result up to order $1/Q$\\ 
for polarized deep-inelastic leptoproduction\\
{\em [Nucl.Phys. B461 (1996) 197]}
}
 
\author{P.J. Mulders$^{1,2}$ and R.D. Tangerman$^1$}
\address{\mbox{}\\
$^1$National Institute for Nuclear Physics and High--Energy
Physics (NIKHEF)\\
P.O. Box 41882, NL-1009 DB Amsterdam, the Netherlands\\
\mbox{}\\
$^2$Department of Physics and Astronomy, Free University \\
De Boelelaan 1081, NL-1081 HV Amsterdam, the Netherlands
}
\maketitle
 
\begin{abstract}
We present the results of the tree-level calculation of deep-inelastic
leptoproduction, including polarization of target hadron and produced
hadron. We also discuss the dependence on transverse momenta of the
quarks, which leads to azimuthal asymmetries for the produced hadrons.
\end{abstract}
 
 
\section{Introduction}
 
In recent years several possible ways to probe the structure of hadrons
in hard scattering processes have been pointed out. Most well-known are the
inclusive lepton-hadron ($\ell H$) scattering experiments that provide
detailed  information on the unpolarized and polarized quark distributions.
This information is valuable as it, within the framework of Quantum 
Chromodynamics (QCD), can be expressed as well-defined matrix elements
of quark and gluon operators within the nucleon. For instance, the
operator product expansion relates moments of quark distributions to
expectation values of local operators that represent static properties
of the nucleon such as its electric or axial charge, baryon number, etc.
Some of these matrix elements can also be determined in an independent
way, such as the axial charge of the nucleon from neutron decay.
We will consider in this paper matrix elements of nonlocal combinations 
of quark fields \cite{Soper-77,Collins-Soper-82}. 
They represent forward quark-target scattering amplitudes that 
can be interpreted as quark momentum distributions or multi-parton
distributions \cite{Jaffe-83}.
 
In inclusive $\ell H$ scattering one can determine combinations of the
quark distributions $f_1^{a}(x)$ and $g_1^{a}(x)$ where $x$ is
the fractional lightcone momentum $x$ = $p^+/P^+$ of a quark
with momentum $p$ in a hadron with momentum $P$. The index $a$ refers to
the flavor of the quark. The antiquark distributions found in inclusive
$\ell H$ scattering are denoted $f_1^{\bar a}(x)$ or $\bar f_1^{a}(x)$, etc.
These quark distributions appear in the structure
functions, which in the Bjorken limit scale in the variable $\xbj$ =
$Q^2/2P\cdot q$, where $Q^2$ = $-q^2$ with $q$ the (large) momentum of the
virtual photon. Restricting ourselves to the exchange of photons, the
experimentally accessible leading structure functions are
\ba
&&\frac{F_2(\xbj,Q^2)}{\xbj} = 2\,F_1(\xbj,Q^2) =
\sum_a e_a^2\,\left( f_1^{a}(\xbj) + \bar f_1^{a}(\xbj) \right), \\
&&2\,\mbox{\Large {\em g}}_1(\xbj,Q^2) =
\sum_a e_a^2\,\left( g_1^{a}(\xbj) + \bar g_1^{a}(\xbj) \right), 
\ea
where $e_a$ are the quark charges.
The functions appear in leading order in $1/Q$ and are referred to as 
twist-two functions since their moments are matrix elements of local 
operators of twist two. 
In naming the quark distributions we extend the scheme proposed
by Jaffe and Ji \cite{Jaffe-Ji-91b,Jaffe-Ji-92}. For the 
unpolarized quark distributions, often denoted as
$f^a(x)$ or $a(x)$, the notation becomes $f_1^{a}(x)$. 
For the polarized quark distributions,  for which one finds $\Delta f^a(x)$ 
or $\Delta a(x)$, the notation becomes $g_1^{a}(x)$, 
 
The only other distribution function that can be obtained in inclusive
deep-inelastic $\ell H$ scattering up to order $1/Q$ is the twist-three
function $g_T^{a}(x)$. 
It appears in the structure function
\be
2\,\mbox{\Large {\em g}}_2(\xbj,Q^2) =
\sum_a e_a^2\,\left( g_2^{a}(\xbj) + \bar g_2^{a}(\xbj) \right), 
\ee
where $g_2^{a}(x)$ = $-g_1^{a}(x)$ + $g_T^{a}(x)$. 
The functions $g_T^{a}$
can be expressed as matrix elements of quark and gluon field operators
\cite{Efremov-Teryaev-84,Ratcliffe-86} and their moments involve 
local operators of twist three \cite{Jaffe-Ji-91a}.
 
In processes involving at least two hadrons, it is possible to measure one
more twist-two quark distribution function. This function, the transverse
spin distribution $h_1^{a}(x)$ (also known as $\Delta_T a(x)$)
and the antiquark distribution $\bar h_1^{a}(x)$ are chirally odd, hence 
they must be combined with some other chirally odd 
structure \cite{Artru-Mekhfi-90,Artru-91}. 
This is possible in Drell-Yan scattering or in semi-inclusive $\ell H$
scattering. In the latter case the chirally odd structure is found in the
quark $\rightarrow$ hadron fragmentation process. At the twist-three level,
two more chirally odd quark distributions, $e^{a}(x)$ and $h_L^{a}(x)$
appear \cite{Jaffe-Ji-91b,Jaffe-Ji-92}.
 
In processes involving at least two hadrons or two jets, one
also has the possibility to investigate azimuthal dependences. This involves
a momentum that is transverse to the large momentum flow, e.g., the azimuthal
dependence of hadrons around the (spacelike) direction defined by the
virtual photon. This is the specific case investigated in this paper. Other
examples are the azimuthal dependence of the produced muon pair in Drell-Yan
scattering or the azimuthal dependence of hadrons around the jet-jet axis in
two-jet events in $e^+e^-$ annihilation.
 
As has been discussed first by Ralston and Soper \cite{Ralston-Soper-79} 
and recently by us \cite{Tangerman-Mulders-95a}, one has
the possibility to measure in total six quark distribution functions depending
on the lightcone fraction $x$ and the transverse momentum squared 
$\bpt^2$, which  depends on $x$ and the quark virtuality $p^2$. 
In this paper we investigate the full tree-level result for
semi-inclusive $\ell H$ scattering up to order $1/Q$ which in addition
to the six twist-two distribution functions involves eight twist-three
functions. After $\bpt$-integration only the three twist-two and three
twist-three ($x$-dependent) functions discussed above remain.
For semi-inclusive scattering one also needs to study
the quark fragmentation process involving a number of twist-two and
twist-three fragmentation functions. The treatment of the fragmentation
functions proceeds along the same lines as the distribution functions with
the exception that one also can have some time-reversal odd structure. This
structure emerges as the outgoing hadron in $\ell H$ scattering is not a
plane wave state in contrast to the incoming hadron.
 
An important motivation to study quark and gluon distribution and fragmentation
functions in hard scattering processes are the existence of factorization
theorems in QCD that imply the universality of these
functions, notwithstanding their (logarithmic) scale dependence. Such 
factorization theorems exist for the $p_\st$-integrated 
functions \cite{Ellis-et-al-79}. They do not exist for the unintegrated 
distributions \cite{Collins-Soper-Sterman-83,Collins-93a}.
Nevertheless, we consider the present complete tree-level result up
to (subleading) order $1/Q$ as an important step. 
It represents the extension of the naive 
parton model result to subleading order and shows the dominant structures 
to be expected in the cross section although QCD corrections, such as 
Sudakov effects, may affect the asymmetries \cite{Dokshitzer-80}. 
At the tree-level the only 
QCD input at order $1/Q$ is the use of the equations of motion 
which assures the electromagnetic gauge invariance.
In view of possible future measurements with electrons or muons at DESY,
CERN or Fermilab it is important to be aware of possible azimuthal
asymmetries, not only because they contain valuable new information on
the quark and gluon structure of hadrons but also in order to estimate
corrections to $p_\st$-integrated quark distributions 
coming from variations in angular acceptance.

The structure of this paper is as follows. In section 2 we present 
the formalism of $\ell H$ scattering followed in section 3 by the
analysis of quark distribution and fragmentation functions. This is
followed by the full tree-level result up to order $1/Q$ in section 4, the
basic result of this paper. The observable consequences of this result
are outlined in the next two sections, starting with the full 
$p_\st$-integrated result for $\ell H$ $\rightarrow$ $hX$ 
in section 5.
The cross sections are given including polarization of lepton and 
target hadron and including measured final state polarization, all of
this for the case of spin 1/2 particles.
In section 6, the result for azimuthal asymmetries involving unintegrated
distribution and fragmentation functions is presented, 
including only polarizations of initial state particles. 
Finally the results are summarized in section 7.

\section{Formalism}

We consider deep-inelastic semi-inclusive lepton-hadron scattering, 
$\ell + H \rightarrow \ell^\prime + h + X$, 
in which a lepton with momentum $l$ scatters off a hadron $H$ with momentum
$P$ and one hadron $h$ with momentum $P_h$ belonging to the current jet
is measured in coincidence
with the scattered lepton with momentum $l^\prime$. The momentum of the
exchanged virtual photon $q$ = $l-l^\prime$ is spacelike with $-q^2$ 
$\equiv$ $Q^2$ $\rightarrow$ $\infty$. We will use the invariants 
$\xbj \equiv Q^2/2P\cdot q$ $\approx$ $ -P_h\cdot q/P_h\cdot P$, 
$y \equiv P\cdot q/P\cdot l \approx Q^2/\xbj s$ (where $s$ is the 
invariant mass squared of the photon-hadron system), 
and $z_h \equiv P\cdot P_h/P\cdot q$ $\approx$ $ - 2P_h\cdot q/Q^2$,
the approximate sign indicating relations valid up to order $1/Q^2$ 
corrections. Furthermore, we use a transverse (two-component) vector 
$\bqt$ which is small (i.e., of order $M$). 
In the frame where $h$ and $H$ are collinear it is precisely the transverse 
component of $q$, while in the frame where $P$ and $q$ are collinear it is 
up to $1/Q^2$ corrections equal to $\bqt \approx -\bP_{h\perp}/z_h$. 
Note that we will henceforth systematically neglect $1/Q^2$ corrections
even if we use equal signs.

The cross section for semi-inclusive lepton-hadron scattering is given
by
\be
\frac{d\sigma}{d\xbj\,dy\,dz_h\,d^2\bqt} =
\frac{\pi\,\alpha^2}{2Q^4}\,y\,z_h\,L_{\mu \nu}\, 2M{\cal W}^{\mu \nu}.
\label{contract}
\ee
The leptonic tensor is standard and is for polarized leptons
(neglecting the lepton masses) given by
\be \label{lepton} 
L_{\mu\nu} (l \lambda ; l^\prime \lambda^\prime)
= \delta_{\lambda \lambda^\prime}\left(
2 l_{\mu}l^\prime_{\nu} + 2 l_{\nu}l^\prime_{\mu}
- Q^2 g_{\mu\nu} +2i\lambda\, \epsilon_{\mu\nu\rho\sigma}
q^\rho l^{\sigma} \right),
\ee
with the lepton helicity being $\lambda$ = $\pm 1$.
The hadronic tensor is given by
\ba
2M{\cal W}_{\mu\nu}( q; P S; P_h S_h ) & = & \frac{1}{(2\pi)^4}\sum_X
\int \frac{d^3 P_X}{(2\pi)^3 2P_X^0}
 (2\pi)^4 \delta^4 (q + P - P_X - P_h)\,
\nonumber \\ &&\qquad \qquad \qquad \mbox{} \times
\langle P S |J_\mu (0)|P_X; P_h S_h \rangle
\langle P_X; P_h S_h |J_\nu (0)|P S \rangle.
\label{hadron2}
\ea
The role of the polarization vectors $S$ and $S_h$ is discussed in
Appendix A. For the inclusive process one finds
\be
\frac{d\sigma}{d\xbj \, dy} =
\frac{\pi\,\alpha^2}{Q^4}\,y \,
L_{\mu \nu}\, 2MW^{\mu \nu},
\ee
with the hadronic tensor given by
\begin{eqnarray}
2M\,W_{\mu \nu}(q;P S) & = &
\frac{1}{2\pi}\sum_X 
\int \frac{d^3 P_X}{(2\pi)^3 2P_X^0}
 (2\pi)^4 \delta^4 (q + P - P_X)
\langle P S |J_\mu (0)|P_X\rangle
\langle P_X |J_\nu (0)|P S \rangle  \nonumber \\
& = &
\frac{1}{2\pi} 
\int d^4 x\ e^{iq\cdot x} \
\langle P S |[ J_\mu (x), J_\nu (0)]|P S \rangle . \label{hadron1}
\end{eqnarray}

In order to expand the leptonic and hadronic tensors it is convenient 
to work with vectors orthogonal
to $q$. A spacelike vector is defined by $q^\mu$ and a timelike
vector defined by $\tilde P^\mu \equiv P^\mu - (P\cdot q/q^2)\,q^\mu$,
\begin{eqnarray}
Z^{\mu} & \equiv & -q^\mu, \\
T^{\mu} & \equiv &  -\frac{q^2}{P\cdot q}\,\tilde P^\mu
           \ = \   2\xbj\,P^\mu +q^\mu, 
\end{eqnarray}
These vectors satisfy $T^2 = Q^2$, $Z^2 = - Q^2$. We will consider normalized
vectors $\hat t^\mu = T^\mu/Q$, $\hat z^\mu = Z^\mu/Q$ = 
$-\hat q^\mu = -q^\mu/Q$.
In the space orthogonal to $\hat z$ and $\hat t$ one has the tensors
\ba
&&g_\perp^{\mu\nu}
 \equiv  g^{\mu\nu} +\hat q^\mu \hat q^\nu -
\hat t^\mu \hat t^\nu,  \\
&&\epsilon_\perp^{\mu \nu} \equiv
  \epsilon^{\mu \nu \rho \sigma} \hat t_\rho \hat q_\sigma
= \frac{1}{P\cdot q}\,\epsilon^{\mu \nu \rho \sigma}
P_\rho q_\sigma. 
\ea
One vector in the perpendicular space is $P_{h\perp}^\mu =
g_\perp^{\mu\nu}P_{h\nu}$. We will frequently use the normalized vector
$\hat h^\mu = P_{h\perp}^\mu/\vert \bP_{h\perp}
\vert$.
 
Also the lepton momenta $l$ and $l^\prime = l-q$ can be expanded in $\hat t$,
$\hat z$ and a perpendicular component,
\be
l^\mu = \frac{2-y}{y}\,T^\mu - \frac{1}{2}\,Z^\mu + l_\perp^\mu
\ =\ \frac{Q}{2}\,\hat q^\mu + \frac{(2-y) Q}{2y}\,\hat t^\mu
+ \frac{Q\sqrt{1-y}}{y}\,\hat x^\mu, 
\ee
where $\hat x^\mu$ = $l_\perp^\mu/\vert \bl_\perp\vert$, is a spacelike
unit-vector in the perpendicular direction lying in the (lepton) scattering
plane. The kinematics in the frame where virtual photon and target are
collinear (including target rest frame) is illustrated in Fig. 1.
With the definition of $\hat x$, we have for the leptonic tensor
\ba
L^{\mu \nu}_{(\ell H)} & = &\frac{Q^2}{y^2} \Biggl[
-2 \left( 1 - y + \frac{1}{2}\,y^2 \right) g_\perp^{\mu \nu}
+ 4(1-y) \hat t^\mu \hat t^\nu
\nonumber \\ && \qquad
+ 4(1-y)\left( \hat x^\mu\hat x^\nu +\frac{1}{2}\,g_\perp^{\mu \nu}\right)
+ 2(2 - y)\sqrt{1-y}\,\,\hat t^{\{ \mu}\hat x^{\nu \}}
\nonumber \\ &&\qquad
-i\lambda\,y(2-y)\,\epsilon_\perp^{\mu \nu}
- 2i\lambda\,y\sqrt{1-y}\,\,\hat t^{\,[\mu} 
\epsilon_\perp^{\nu] \rho} \hat x_\rho \Biggr].
\ea

Theoretically, it is often convenient to work in the frame in which the two
hadrons  $H$ and $h$ are collinear. Except for corrections proportional to
$M^2/Q^2$ the vectors $P^\mu/Q$ and $P_h^\mu/Q$ determine two lightlike
directions  $n_+^\mu$ and $n_-^\mu$, respectively.  Using for momenta the
representation  $p$ = $[p^-,p^+,\bpt]$,
 where $p^\pm = (p^0 \pm p^3)/\sqrt{2}$, one obtains in a frame in which
$H$ and $h$ are collinear
\begin{eqnarray}
P & \ = \ & \left[ \frac{\xbj M^2}{A\sqrt{2}},
\frac{A}{\xbj \sqrt{2}}, \bnull_T \right]
 \ \equiv \  \frac{Q}{\xbj\sqrt 2}\,n_+ + \frac{\xbj M^2}{Q\sqrt{2}}\,n_-, 
\label{collinearP} \\
P_h & \ = \ & \left[ \frac{z_h Q^2}{A\sqrt{2}}, 
\frac{A M_h^2}{z_h Q^2 \sqrt{2}},
\bnull_T \right]
\ \equiv \  \frac{z_h\,Q}{\sqrt 2}\,n_- + \frac{M_h^2}{z_h\,Q\sqrt{2}}\,n_+, 
\label{collinearPh} \\
q & \ = \ & \left[ \frac{Q^2}{A\sqrt{2}}, -\frac{A}{\sqrt{2}},
\bqt \right]
\ = \ \frac{Q}{\sqrt 2}\,n_- - \frac{Q}{\sqrt{2}}\,n_+ 
+ q_\st,
\label{collinearq}
\end{eqnarray}
where $A$ fixes the frame and $n_+\cdot n_-$ = 1. 
In a collinear frame as given here, the photon
acquires a transverse momentum $q_\st$ with length 
$q_\st^2$ = $-\bqt^2$ = $-Q_T^2$. It is easy to see that $P_{h\perp}^\mu 
= -z_h q_\st^\mu$ and thus $Q_T$ = $\vert \bP_{h\perp}\vert/z_h$.
Transverse projection operators are given by
\ba
&&g^{\mu\nu}_{\scriptscriptstyle T} \ \equiv \ g^{\mu\nu} 
- n_+^{\,\{\mu} n_-^{\nu\}}, \\
&&\epsilon^{\mu\nu}_{\scriptscriptstyle T} \ \equiv 
\ \epsilon^{\mu\nu\rho\sigma} n_{+\rho}n_{-\sigma},
\ea
where the brackets around the indices indicate symmetrization of these
indices. Note that these {\em transverse} projectors are not identical to
the {\em perpendicular} ones defined above if the transverse momentum 
of the outgoing hadron does not vanish. 
The lightlike directions, however, can easily be expressed in $\hat t$, 
$\hat z$ and a perpendicular component, 
\ba
n_+^\mu & = & \frac{1}{\sqrt{2}} \left[ \hat t^\mu + \hat z^\mu \right], 
\label{transverse1} \\
n_-^\mu & = & \frac{1}{\sqrt{2}} \left[ \hat t^\mu - \hat z^\mu 
- 2\,\frac{q_\st}{Q} \right],
\label{transverse2} 
\ea
showing that the differences are of order $1/Q$. Especially for the treatment
of azimuthal asymmetries, it is important to keep track of these differences.

\section{Quark correlation functions}

In this section we investigate the 'soft' hadronic matrix elements that
appear in the diagrammatic expansion of a hard scattering amplitude
\cite{Ellis-Furmanski-Petronzio-83}.
The most important matrix elements that can be measured in deep-inelastic
semi-inclusive lepton-hadron scattering are of two kinds. The first
\cite{Soper-77,Collins-Soper-82,Jaffe-83} is
\be
\Phi_{ij}(P,S;p) = \frac{1}{(2\pi)^4}\int d^4x\ e^{i\,p\cdot x}
\langle P,S \vert \overline \psi_j(0) {\cal L}(0,x;\mbox{path})\psi_i(x) 
\vert P,S \rangle,
\ee
where a summation over color indices is implicit and flavor indices are
omitted. 
This expression, referred to as a quark-quark correlation function, is a
forward matrix elements of quark fields with explicit  Dirac indices $i$
and $j$ in a hadronic state characterized by its momentum $P$ and spin
vector $S$. The quark fields are accompanied by a  path ordered
exponential (link operator),
\be
{\cal L}(0,x;\mbox{path}) = {\cal P} \exp \left(-ig\int_0^{x} ds^\mu 
\,A_\mu(s)\right),
\ee
necessary to render the definition color gauge-invariant. It introduces,
however, a path-dependence. In order to obtain a simple parton
interpretation (Fig. 2) for the correlation functions the choice of path
is important, as we will see later. Technical details will be discussed in
Appendix B. The other type of matrix elements measured in semi-inclusive
lepton-hadron scattering are the quark decay functions \cite{Collins-Soper-82} 
(Fig. 2)
\begin{eqnarray}
\Delta_{ij}(P_h,S_h;k) & = & \sum_X \frac{1}{(2\pi)^4}\int d^4x\ e^{ik\cdot x}\,
\langle 0 \vert {\cal L}(0,x;\mbox{path})\psi_i(x) \vert P_h, S_h; X \rangle
\langle P_h, S_h;X \vert \overline \psi_j(0) \vert 0 \rangle \nonumber \\
& = & \frac{1}{(2\pi)^4}\int d^4x\ e^{ik\cdot x}\,
\langle 0 \vert {\cal L}(0,x;\mbox{path})\psi_i (x) a_h^\dagger
a_h \overline \psi_j(0) \vert 0 \rangle,
\end{eqnarray}
where an averaging over color indices is implicit. Again a quark link operator 
accompanies the quark fields. The second way of writing using hadron 
creation and annihilation
operators is to be considered as a shorthand of the first expression,
involving (out-)states $\vert P_h, S_h; X\rangle$ where $P_h$ and $S_h$ are
the momentum and spin vectors for the produced hadron.

The correlation functions are constrained by the hermiticity properties of
the fields, and by parity and time reversal invariance.
Without constraints from time reversal invariance one has
\cite{Ralston-Soper-79,Tangerman-Mulders-95a}
\begin{eqnarray}
\Phi(P,S;p) & = &
A_1 + A_2\,{\not\! P} + A_3 {\not\! p}
+ A_{4}\,\sigma^{\mu \nu} P_\mu p_\nu 
+ i\,A_{5}\,(p\cdot S) \gamma_5 ,
+ A_6 \,{\not\! S} \gamma_5
\nonumber \\ & & 
+ A_7\,(p\cdot S) {\not\! P} \gamma_5
+ A_8\,(p\cdot S) {\not\! p} \gamma_5 
+ i\,A_9\,\sigma^{\mu \nu} \gamma_5 \,S_\mu P_\nu
+ i\,A_{10}\,\sigma^{\mu \nu} \gamma_5 \,S_\mu p_\nu
\nonumber \\ & & 
+ i\,A_{11}\,(p\cdot S)\,\sigma^{\mu \nu} \gamma_5 \,p_\mu P_\nu
+ A_{12}\, \epsilon_{\mu \nu \rho \sigma}\gamma^\mu P^\nu p^\rho S^\sigma,
\label{expansion}
\end{eqnarray}
where the first four terms do not involve
the hadron polarization vector. Hermiticity requires all the amplitudes
$A_i$ = $A_i(p\cdot P, p^2)$ to be real. The amplitudes $A_4$, $A_5$
and $A_{12}$ vanish when also time reversal invariance applies. A similar
decomposition in 12 amplitudes can be made for $\Delta$, but because
of the fact that out-states appear in the definition of the correlation
functions time reversal invariance cannot be used and none of the amplitudes 
vanish. Because the time-reversal condition for $A_4$, $A_5$ and $A_{12}$ reads
$A_i^\ast$ = $-A_i$ we refer to these amplitudes as time-reversal odd.

The above matrix elements are assumed to be nonzero only for quark momenta
limited to a characteristic hadronic scale, which is much smaller than
$Q^2$. This means that in the above matrix elements 
$p^2$, $k^2$, $p\cdot P$ and $k\cdot P_h$ $\ll$ $Q^2$.
With the definition of momenta in the collinear frame (Eqs \ref{collinearP}
- \ref{collinearq}) one then arrives at the Sudakov decomposition,
for the quark momenta,
\ba
p & = & \left[ \frac{(p^2+\bpt^2)}{2x\,P^+},
\ x\,P^+,\ \bpt\right]
\ \equiv \ \frac{x\,Q}{\xbj\sqrt{2}}\,n_+ 
+ \frac{\xbj (p^2+\bpt^2)}{x\,Q\sqrt{2}}
\,n_- + p_\st
\ \approx \ x\,P + p_\st,
\\
k & = & \left[ \frac{P_h^-}{z},
\ \frac{z\,(k^2+\bkt^2)}{2\,P_h^-},\ \bkt\right]
\ \equiv \ \frac{z_h\,Q}{z\sqrt{2}}\,n_- 
+ \frac{z\,(k^2+\bkt^2)}{z_h\,Q\sqrt{2}} \,n_+ 
+ k_\st
\ \approx \ \frac{1}{z}\,P_h + k_\st,
\ea
where in the last expressions for $p$ ($k$) the terms proportional to
$n_-$ ($n_+$), irrelevant in leading, ${\cal O}(1)$, or subleading,
${\cal O}(1/Q)$, order have been omitted. It turns out that up to 
subleading order only the combinations $\int dp^-\,\Phi(P,S;p)$ and $\int
dk^+\,\Delta(P_h,S_h;k)$ are important. For cross sections integrated
over transverse momenta
only the combinations $\int dp^-d^2\bpt\,\Phi(P,S;p)$ and $\int
dk^+d^2\bkt\,\Delta(P_h,S_h;k)$ are of importance at leading order.  

The functions that appear in hard scattering processes can be
expressed as specific Dirac projections of the correlation functions, 
integrated over $p^-$, 
\begin{eqnarray}
\Phi^{[\Gamma]}(x,\bpt) & = &
\left. \frac{1}{2}\int dp^-\ Tr(\Phi\,\Gamma) \right|_{p^+ =
x P^+,\ \bpt} \nonumber \\
& = & \left. \int \frac{d\xi^-d^2\xi_\st}{2\,(2\pi)^3} 
\ e^{ip\cdot \xi}
\,\langle P,S \vert \overline \psi (0)\,\Gamma\,{\cal L}(0,\xi;n_-)
\,\psi(\xi) \vert P,S \rangle \right|_{\xi^+ = 0}. \label{projection}
\end{eqnarray}
In general the matrix element contains an infinite tower of operators
because of the link operator. Choosing the path in the
definition of $\Phi$ to lie in the plane $x^+$ = 0 and in essence along the
direction $x^-$, indicated with 'path' = $n_-$ (see Appendix B), one finds 
that in an appropriately chosen gauge in leading order only one (bilocal) 
operator combination contributes, which allows a parton interpretation. 

The projections $\Phi^{[\Gamma]}$ depend on the fractional momentum
$x$ = $p^+/P^+$ and on $\bpt$ and furthermore on the hadron momentum
$P$ (in essence only $P^+$ and $M$).
Depending on the Lorentz structure of the Dirac matrix $\Gamma$ the
projections $\Phi^{[\Gamma]}$ can be ordered according to powers
of $M/P^+$ multiplied with a function depending only on $x$ and $\bpt^2$ = 
$-p_\st^2$.
It is easy to convince oneself that each factor $M/P^+$ 
leads to a suppression with a power $M/Q$ in cross sections 
\cite{Levelt-Mulders-94a}. In
analogy with inclusive scattering we therefore refer to the projections
as having a 'twist' $t$ related to the power $(M/P^+)^{t-2}$ that
appears. With this definition the moments (in $x$) of the $\bpt$-integrated 
functions indeed involve local operators of twist $t$ 
\cite{Collins-Soper-82}.

It is convenient to use for the spin vector $S$ also a Sudakov
decomposition,
\be 
S = \left[-\frac{\lambda M}{2P^+},\ \frac{\lambda
P^+}{M},\ \bS_\st\right] \ = \ \frac{\lambda\,Q}{M\xbj\sqrt{2}}\,n_+
- \frac{\lambda\,M\xbj}{Q\sqrt{2}}\,n_- + S_T
\ \approx \ \lambda\,\frac{P}{M} + S_T,
\ee
with $\lambda^2 + \bS_\st^2$ = 1 and again neglecting the contribution
(proportional to $n_-$) which is irrelevant up to subleading order.
The following distribution functions appear at leading order in
$(M/P^+)^0$ (twist-two), where as said before we omit flavor indices,
\begin{eqnarray}
& & \Phi^{[\gamma^+]}(x,\bpt) =
f_1(x ,\bpt^2) ,
\\ & & \Phi^{[\gamma^+ \gamma_5]}(x,\bpt) =
\lambda\,g_{1L}(x ,\bpt^2)  
+ g_{1T}(x ,\bpt^2)\,\frac{(\bpt\cdot\bS_\st)}{M} ,
\label{chirality}
\\ & & \Phi^{[ i \sigma^{i+} \gamma_5]}(x,\bpt) =
S_T^i\,h_{1T}(x ,\bpt^2)
+ \frac{p_\st^i}{M}\left[ \lambda\,h_{1L}^\perp(x ,\bpt^2)
+ h_{1T}^\perp(x ,\bpt^2)\,\frac{(\bpt\cdot\bS_\st)}{M}\right]\nonumber \\
& & \qquad \qquad
= S_T^i\,h_1(x ,\bpt^2)
+ \frac{\lambda p_\st^i}{M} \,h_{1L}^\perp(x ,\bpt^2)
+ \frac{\left(p_\st^i p_\st^j 
- \frac{1}{2}\bpt^2\delta_{ij}\right) S_T^j}{M^2}
\,h_{1T}^\perp(x ,\bpt^2) , 
\label{transversity}
\end{eqnarray}
with $h_1$ = $h_{1T} + (\bpt^2/2M^2)\,h_{1T}^\perp$.
At subleading order $(M/P^+)^1$ (twist-three) one finds
\begin{eqnarray}
& & \Phi^{[1]}(x,\bpt) =
\frac{M}{P^+}\,e(x ,\bpt^2)
\\ & & \Phi^{[\gamma^i]}(x,\bpt) =
\frac{p_\st^i}{P^+}\,f^\perp(x ,\bpt^2),
\\ & & \Phi^{[ \gamma^i \gamma_5]}(x,\bpt) =
\frac{M\,S_T^i}{P^+} 
\, g_T^\prime(x ,\bpt^2)
+ \frac{p_\st^i}{P^+}
\left[ \lambda\,g_L^\perp(x ,\bpt^2)\,
+ g_T^\perp(x ,\bpt^2)\,\frac{(\bpt\cdot\bS_\st)}{M} \right] \nonumber \\
& & \qquad \quad
= \frac{M\,S_T^i}{P^+} \, 
g_T(x ,\bpt^2)
+ \frac{\lambda p_\st^i}{P^+}\,g_L^\perp(x ,\bpt^2)
+ \frac{\left(p_\st^i p_\st^j 
- \frac{1}{2}\bpt^2\delta_{ij}\right) S_T^j}{M^2}
\,g_T^\perp(x ,\bpt^2),
\\ & & \Phi^{[i \sigma^{ij} \gamma_5]}(x,\bpt) =
\frac{S_T^ip_\st^j
-p_\st^iS_T^j}{P^+}
\,h_T^\perp(x ,\bpt^2)
\\ & & \Phi^{[ i\sigma^{+-}\gamma_5 ]}(x,\bpt) =
\frac{M}{P^+} \left[ \lambda\,h_L(x ,\bpt^2)
+ h_T(x ,\bpt^2)\,\frac{(\bpt\cdot\bS_\st)}{M} \right],
\end{eqnarray}
with $g_T$ = $g_T^\prime + (\bpt^2/2M^2)\,g_T^\perp$. From 
here on we will often use the shorthand notation
\be
g_{1s}(x, \bpt) \equiv
\lambda\,g_{1L}(x ,\bpt^2)
+ \frac{(\bpt\cdot\bS_\st)}{M}\,g_{1T}(x ,\bpt^2) ,
\ee
and similarly shorthand notations $h^\perp_{1s}$, $g_s^\perp$ and
$h_s$. The results after integration over $\bpt$ are lightcone 
correlation functions $\Phi^{[\Gamma]}(x)$ depending only on one lightcone
momentum component $p^+$ = $x\,P^+$,
for which the nonlocality is restricted to $x^-$, i.e., $x^+ = 0$ and 
$\bx_T = \bnull_T$. 
Only the functions $f_1$, $g_1$ =
$g_{1L}$, $h_1$, $e$, $g_T$ and $h_L$ are nonvanishing
upon integration over $\bpt$. Upon further integration over $x$ a local
matrix element is found, e.g., relating the valence distributions to the
quark numbers, $\int dx\,(f_1^a(x) - \bar f_1^a(x))$ = $\langle P,S
\vert \overline \psi^a(0)\gamma^+\psi^a(0)\vert P,S\rangle/2P^+$
= $n_a - \bar n_a$.

In naming the functions, we have extended the scheme introduced by Jaffe and Ji.
All functions obtained after tracing with a scalar ($1$) or pseudoscalar
($i\gamma_5$) Dirac matrix are given the name $e_{..}$, those traced with 
a vector matrix ($\gamma^\mu$) are given the name $f_{..}$, those traced
with an axial vector matrix ($\gamma^\mu\gamma_5$) are given the name $g_{..}$ 
and finally those traced with the second rank tensor ($i\sigma^{\mu\nu}
\gamma_5$) are given the name $h_{..}$. A subscript $'1'$ is given to the
twist-two functions, subscripts $'L'$ or $'T'$ refer to the connection with the
hadron spin being longitudinal or transverse and a superscript $'\perp'$ 
signals the explicit presence of transverse momenta with a noncontracted
index.

The twist-two distribution functions have a natural interpretation as parton
densities. Using the good and bad components of the quark fields,
$\psi_{\pm} = P_{\pm}\psi$ with $P_{\pm}$ = $\gamma^-\gamma^+/2$ one finds
that the operator combination in $\Phi^{[\gamma^+]}$ is just a density,
$\psi_+^\dagger\,\psi_+$.
The other twist-two correlations functions involve in addition chiral
projections of the quark field, $\psi_{\scriptstyle R/L}$ =
$P_{\scriptstyle R/L}\psi$ with $P_{\scriptstyle R/L}$ =
$(1\pm\gamma_5)/2$ or transverse spin projections
$\psi_{_{\uparrow/\downarrow}}$ = $P_{_{\uparrow/\downarrow}}\psi$ with
$P_{_{\uparrow/\downarrow}}$ = $(1\pm\gamma^1\gamma_5)/2$. The correlation
functions then become,
\ba
&& \Phi^{[\gamma^+]} = f_1 
= f_1^{\scriptstyle (R)} + f_1^{\scriptstyle (L)}
= f_1^{^{(\uparrow)}} + f_1^{^{(\downarrow)}}, \\
&& \Phi^{[\gamma^+\gamma_5]}  
= f_1^{\scriptstyle (R)} - f_1^{\scriptstyle (L)}, \\
&& \Phi^{[i\sigma^{1+}\gamma_5]}  
= f_1^{^{(\uparrow)}} - f_1^{^{(\downarrow)}},
\ea
where $f_1^{\scriptstyle (R)}$ is a matrix element involving the fields 
$\psi_{_{+R}}^\dagger\psi_{_{+R}}$, etc. Thus, the correlation function
$\Phi^{[\gamma^+]}(x,\bpt)$
is just the unpolarized quark distribution, which integrated over $\bpt$
gives the familiar lightcone momentum distribution $\Phi^{[\gamma^+]}(x)$
= $f_1(x)$. 
The correlation function $\Phi^{[\gamma^+\gamma_5]}(x,\bpt)$
is the chirality distribution (for massless quarks helicity distribution).
It can only be measured in a polarized target, as is evident from the
presence of $\lambda$ and $\bS_\st$ in Eq.~\ref{chirality}. In a transversely
polarized (spin 1/2) target the chirality distribution is proportional to
the transverse momentum of the quark along $\bS_\st$. Integrated over $\bpt$
only one distribution survives in the correlation function,
$\Phi^{[\gamma^+\gamma_5]}(x)$ = $\lambda\,g_1(x)$, which 
can only be measured in a longitudinally polarized target.
The correlation function $\Phi^{[i\sigma^{1+}\gamma_5]}(x,\bpt)$ is the
transverse spin distribution and can also only be measured in a polarized
target. The possible dependence on target spin and transverse momenta
involves three functions, as can be seen in Eq.~\ref{transversity}. Using
the second expression one sees that one term (involving $h_1$) is insensitive 
to the direction of the quark transverse momentum; this part of the
correlation function can only be measured in a transversely polarized target.
The other two terms in Eq.~\ref{transversity} depend on rank 1 and rank 2
combinations of the transverse momentum of the quark and are accessible
in longitudinally and transversely polarized targets, respectively.  
Integrated over $\bpt$ only one transverse spin distribution survives,
$\Phi^{[i\sigma^{i+}\gamma_5]}(x)$ = $S_T^i\,h_1(x)$, which can only be 
measured in a transversely polarized target. It involves a 
{\em chirally odd} \cite{Jaffe-Ji-92} operator structure, which in a
cross section can only appear in combination with
another chirally odd operator such as a quark mass term in inclusive 
scattering, a chirally odd fragmentation function in semi-inclusive
scattering or a chirally odd antiquark distribution in Drell-Yan scattering.

The twist-three distributions cannot be expressed as densities or differences
of densities. They are related to matrix elements containing one good and
one bad quark field. The bad field can in principle be replaced by a good field
and a transverse gluon field using the equations of motion.
Up to ${\cal O}(1/Q)$ one anyway needs to include quark-quark-gluon matrix
elements. For this one needs to consider
\begin{eqnarray}
& & \Phi^\alpha_{D\,ij}(P,S;p) = \frac{1}{(2\pi)^4}\int d^4x\ e^{i\,p\cdot x}
\langle P,S \vert \overline \psi_j(0)\,{\cal L}(0,x;n_-)
\,iD^\alpha (x) \psi_i(x) \vert P,S \rangle. \\
& & \Phi^{\alpha \dagger}_{D\ ij}(P,S;p) = \frac{1}{(2\pi)^4}
\int d^4x\ e^{i\,p\cdot x} \langle P,S \vert
\overline \psi_j(0) \,iD^\alpha (0) \,{\cal L}(0,x;n_-)
\psi_i(x) \vert P,S \rangle,
\end{eqnarray}
where $iD^\alpha$ = $i\partial^\alpha +gA^\alpha$.
The twist analysis is performed by considering the projections
\begin{eqnarray}
\Phi_D^{\alpha [\Gamma ]}(x,\bpt) & = &
\left. \frac{1}{2} \int dp^- \ Tr \left( \Phi_D^\alpha \Gamma \right)
\right|_{p^+ \,=\,x P^+,\ \bpt} \nonumber \\ & = &
\left. \int \frac{d\xi^- d^2\xi_\perp}{2\,(2\pi)^3} \ e^{i\,p\cdot \xi}
\langle P,S \vert \overline \psi(0) \Gamma \,{\cal L}(0,\xi;n_-)
\,iD^\alpha (\xi) \psi(\xi) \vert P,S \rangle \right|_{\xi^+ \,=\, 0} ,
\end{eqnarray}
where ${\cal L}$ is the same link operator as the one used for the quark-quark
correlation functions.
Because of the choice of link, which always lies along the minus direction 
(except for the points $\xi^-$ = $\pm \infty$), one has for the
correlation function with a longitudinal gluon field the relation 
\be
\left. \int \frac{d\xi^- d^2\xi_\perp}{2\,(2\pi)^3} \ e^{i\,p\cdot \xi}\,
\langle P,S \vert \overline \psi_j(0)\Gamma {\cal L}(0,\xi;n_-)
\,iD^+\psi_i(\xi) \vert P,S \rangle \right|_{\xi^+ \,=\, 0} 
= p^+ \,\Phi^{[\Gamma]}(x,\bpt).
\label{eq45}
\ee
The projections obtained for the quark-quark-gluon correlation functions
with transverse gluon fields are not all independent from the
ones defined for the quark-quark correlation functions, either. Some can be
connected to quark-quark correlation functions with one good and one bad
quark field using the QCD equation of motion, $(i{\not\! D} - m)\psi(x)$ = 0.
This gives the relations
\begin{eqnarray}
& & g_{{\scriptscriptstyle T}\alpha\beta}\,
\Phi_{D}^{\alpha [\sigma^{\beta +} ]} = 
\epsilon_{{\scriptscriptstyle T}\alpha \beta}
\, \Phi_{D}^{\alpha [i\sigma^{\beta+}\gamma_5]}
= i\left( Mx\,e - m\,f_1\right) 
+ \epsilon_{{\scriptscriptstyle T}\,ij}\,p_T^i S_T^j \,
xh_T^\perp,
\label{eq46}
\\ & & g_{{\scriptscriptstyle T}\alpha\beta}\,
\Phi_{D}^{\alpha [i\sigma^{\beta +}\gamma_5 ]}
= M x \,h_s - m\,g_{1s},
\\ & &
g_{\scriptscriptstyle T}^{\alpha\beta}\,
\Phi_{D\,\beta}^{[\gamma^{+}]} 
- i\epsilon_{\scriptscriptstyle T}^{\alpha \beta}\,
\Phi_{D\,\beta}^{\ [\gamma^{+}\gamma_5]}
= p_T^\alpha\,xf^\perp
- i\epsilon_{\scriptscriptstyle T}^{\alpha \beta}p_{T\beta}
\left( xg_s^\perp - \frac{m}{M}\,h_{1s}^\perp \right)
- i\epsilon_{\scriptscriptstyle T}^{\alpha \beta}S_{T\beta} 
\,\left( Mx\,g_T^\prime - m\, h_{1T} \right).
\end{eqnarray}
A useful quantity to consider is the (color gauge-invariant) correlation
function
\be
\Phi_{A}^{\alpha [\Gamma]}(x,\bpt) 
\equiv  \Phi_D^{\alpha [\Gamma]}(x,\bpt) 
- p^\alpha\,\Phi^{[\Gamma]}(x,\bpt).
\ee
{}From Eq.~\ref{eq45} one sees that $\Phi_A^{+ [\Gamma]}$ = 0, 
while for the transverse indices $\alpha$ = 1 or 2 it reduces after gauge 
fixing to the pure quark-quark-gluon matrix element (see Fig. 3),
\be
\Phi_{A}^{\alpha [\Gamma ]}(x,\bpt)  \Rightarrow
g\left. \int \frac{d\xi^- d^2\xi_\perp}{2\,(2\pi)^3} \ e^{i\,k\cdot \xi}
\langle P,S \vert \overline \psi(0) \Gamma 
\,A^\alpha (\xi) \psi(\xi) \vert P,S \rangle \right|_{\xi^+ \,=\, 0} ,
\label{eq50}
\ee
needed to calculate the leptoproduction cross sections in subleading
order. We use the correlation function $\Phi_A^\alpha$ to identify {\em 
interaction-dependent} combinations in the distribution functions. 
As an example, from the first relation above (Eq.~\ref{eq46}) we obtain
\be
\epsilon_{{\scriptscriptstyle T}\alpha \beta}
\, \Phi_{A}^{\alpha [i\sigma^{\beta+}\gamma_5]}
= i\left( Mx\,e - m\,f_1\right) 
- \epsilon_{{\scriptscriptstyle T}\,ij}\,p_\st^i S_T^j \,
\left( h_{1T} - xh_T^\perp\right) \equiv i\,Mx\,\tilde e
+ \epsilon_{{\scriptscriptstyle T}\,ij}\,p_\st^i S_T^j \,x\tilde h_T^\perp.
\ee
In this way we can rewrite all twist-three functions in a part containing
twist-two distribution functions and an interaction-dependent part,
e.g., $e = (m/Mx)\,f_1 + \tilde e$.
The results for all twist-three functions are explicitly given in
Appendix C, including results for the $\bpt$-integrated functions.

The antiquark distribution functions in a hadron are obtained from the 
matrix elements
\be
\overline \Phi_{ij}(P,S;p) =
\frac{1}{(2\pi)^4}\int d^4x\ e^{-i\,p\cdot x}
\langle P,S \vert {\cal L}(0,x;n_-)\psi_i(x) 
\overline \psi_j(0) \vert P,S \rangle.
\ee
The antiquark distributions should be defined consistent with the replacement
$\psi \rightarrow \psi^c = C\overline\psi^T$, or $\overline \Phi^{[\Gamma]}$
= $+ \Phi^{c[\Gamma]}$ for $\Gamma$ = $\gamma_\mu$, $i\sigma_{\mu\nu}\gamma_5$
and $i\gamma_5$ and $\overline \Phi^{[\Gamma]}$ 
= $- \Phi^{c[\Gamma]}$ for $\Gamma$ = $1$ and $\gamma_\mu\gamma_5$. Finally,
the anticommutation relations for fermions can be used to obtain,
within the (connected) matrix elements, the symmetry relation
\be
\overline \Phi_{ij}(P,S;k) = - \Phi_{ij}(P,S;-k).
\ee
For the distribution functions this gives the symmetry relations
$\bar f_1(x,\bpt^2) = - f_1(-x,\bpt^2)$
and identically for $g_{1T}$, $h_{1T}$, $h_{1T}^\perp$,
$g_L^\perp$ and $h_L$, while
$\bar g_{1L}(x,\bpt^2) = g_{1L}(-x,\bpt^2)$
and identically for $h_{1L}^\perp$, $e$, $f^\perp$, $g_T^\prime$,
$g_T^\perp$, $h_T^\perp$ and $h_T$.

The procedure for the fragmentation functions obtained from the matrix
element $\Delta(P_h,S_h;k)$ is similar. One again starts with the general
decomposition in twelve amplitudes $A_i(k^2,k\cdot P_h)$, replacing $P
\rightarrow P_h$ and $S \rightarrow S_h$. In this case time reversal 
invariance cannot be applied, as the matrix elements $\Delta_{ij}$ 
involve out-states
containing a hadron $h$, in contrast to the plane wave states in the
case of the matrix elements $\Phi_{ij}$. A consequence is that there are more
fragmentation functions. 
The functions that appear in hard scattering processes can again
be expressed as specific Dirac projections of the correlation functions,
integrated over $k^+$,
\begin{eqnarray}
\Delta^{[\Gamma]}(z,\bkt^\prime) & = &
\left. \frac{1}{4z}\int dk^+\ Tr(\Delta\,\Gamma)\right|_{k^- =
P_h^-/z,\ \bkt} \nonumber \\
& = & \left. \int \frac{d\xi^+d^2\xi_T}{4z\,(2\pi)^3} \
e^{ik\cdot \xi} \,Tr  \langle 0 \vert {\cal L}(0,\xi;n_+)
\,\psi (\xi) a_h^\dagger
a_h \overline \psi(0) \Gamma \vert 0 \rangle \right|_{\xi^- = 0} .
\end{eqnarray}
The arguments of the functions are the lightcone fraction $z$ = $P_h^-/k^-$ and
the transverse momentum $\bkt^\prime$ = $-z\bkt$, which is the perpendicular 
momentum of the hadron $h$ with respect to the quark momentum.
In the definition of the correlation functions $\Delta$ the path is chosen
in the plane $x^- = 0$, in essence along the $x^+$-direction, indicated
with 'path' = $n_+$ (see Appendix B).
The spin vector is now parametrized as
\be
S_h = \left[ \frac{\lambda_h\,P_h^-}{M_h}, - \frac{\lambda_h\,M}{2P_h^-},
\bS_{hT} \right] =
\frac{\lambda_h\,zQ}{M_h\sqrt{2}}\,n_- + \frac{\lambda_h\,M_h}{zQ\sqrt{2}}\,n_+
+S_{hT} \approx \lambda_h\,\frac{P_h}{M_h} + S_{hT}.
\label{spinh}\ee
In the last expression the contribution
proportional to $n_+$, which is irrelevant up to subleading order is omitted.
The role of the final state spin vector in determining the polarization
of the produced hadrons is explained in Appendix A.
The following fragmentation functions appear at leading order $(M_h/P_h^-)^0$
(twist-two),
\begin{eqnarray}
& & \Delta^{[\gamma^-]}(z,\bkt^\prime) =
\hF_1(z,\bkt^{\prime 2}) + \frac{\epsilon_{{\scriptscriptstyle T}\,ij}
\,k_T^i S_{hT}^j}{M_h}\,\hF_{1T}^\perp(z,\bkt^{\prime 2}) ,
\\ & & \Delta^{[\gamma^-\gamma_5]}(z,\bkt^\prime) =
\hG_{1s}(z,\bkt^\prime)
\\ & & \Delta^{[i \sigma^{i-} \gamma_5]}(z,\bkt^\prime) =
S_{hT}^i\,\hH_{1T}(z,\bkt^{\prime 2})
+ \frac{k_T^i}{M_h}\,\hH_{1s}^\perp(z,\bkt^\prime)
+ \frac{\epsilon_{\scriptscriptstyle T}^{ij} k_{T j}}{M_h}
\,\hH_1^\perp(z,\bkt^{\prime 2}).
\end{eqnarray}
At subleading order $(M_h/P_h^-)^1$ (twist-three) one finds
\begin{eqnarray}
& & \Delta^{[1]}(z,\bkt^\prime) =
\frac{M_h}{P_h^-}\,\hE(z,\bkt^{\prime 2}) ,
\\ & & \Delta^{[\gamma^i]}(z,\bkt^\prime) =
\frac{k_T^i}{P_h^-}\,\hF^\perp(z,\bkt^{\prime 2}),
\\ & & \Delta^{[i\gamma_5]}(z,\bkt^\prime) =
\frac{M_h}{P_h^-} \,\hE_s(z,\bkt^\prime),
\\ & & \Delta^{[\gamma^i \gamma_5]}(z,\bkt^\prime) =
\frac{M_h\,S_{hT}^i}{P_h^-} \, \hG_T^\prime(z,\bkt^{\prime 2})
+ \frac{k_T^i}{P_h^-}\,\hG_s^\perp(z,\bkt^\prime),
\\ & & \Delta^{[ i \sigma^{ij} \gamma_5]}(z,\bkt^\prime) =
\frac{S_{hT}^i k_T^j-k_T^i S_{hT}^j}{P_h^-}\,\hH_T^\perp(z,\bkt^{\prime 2})
+\frac{M_h\,\epsilon_{\scriptscriptstyle T}^{ij}}{P_h^-}\,\hH(z,\bkt^{\prime 2}),
\\ & & \Delta^{[i\sigma^{-+}\gamma_5]}(z,\bkt^\prime) =
\frac{M_h}{P_h^-} \,\hH_s(z,\bkt^\prime).
\end{eqnarray}
In the above expressions we have used the shorthand notations $\hG_{1s}$ etc. 
standing for
\be
\hG_{1s}(z,\bkt^\prime) = \lambda_h\,\hG_{1L}(z,\bkt^{\prime 2})
+ \frac{(\bkt\cdot\bS_{hT})}{M_h}\,\hG_{1T}(z,\bkt^{\prime 2}).
\ee
There are more functions allowed in this case because of the non-applicability
of time-reversal invariance, allowing more amplitudes in the expansion
for $\Delta_{ij}$ as compared to that for $\Phi_{ij}$ (see discussion
following Eq.~\ref{expansion}). The fragmentation functions which have
no equivalent in the distribution functions, $\hF_{1T}^\perp$, $\hH_1^\perp$,
$\hE_L$, $\hE_T$ and $\hH$, will be referred to as {\em time-reversal odd}.
The results after integration over $\bkt^\prime$ define the lightcone 
correlation functions $\Delta^{[\Gamma]}(z)$.
Nonvanishing upon $\bkt^\prime$-integration are the functions 
$\hF_1$, $\hG_1$ = $\hG_{1L}$,
$\hH_1$ = $\hH_{1T} + (\bkt^2/2M_h^2)\,\hH_{1T}^\perp$, $\hE$, $\hE_L$,
$\hG_T$ = $\hG_T^\prime + (\bkt^2/2M_h^2)\,\hG_T^\perp$, $\hH$ and 
$\hH_L$ and $\hF_{1T}^{\perp (1)}$ = $(\bkt^2/2M_h^2)\,\hF_{1T}^{\perp}$.
Summing over all produced hadrons $h$ and integrating over $z$ one can use
completeness in the final state to see e.g.\ that one has (for any flavor) the
momentum sum rule $\sum_h\int dz\,z\hF_1(z) = 1$.

A note on the names of the fragmentation functions. 
Because of the analogy with the distribution functions in essence the
same names are used but using the capital letter, except for the functions 
corresponding to the $f_{..}$ distribution functions, for which we have used
the more familiar name $D_{..}$. The connection with the
naming scheme used by Jaffe and Ji, namely using the same letter as the
distribution functions with a hat is straightforward. We have used capital
letters in order to avoid double symbols over the function names.

The twist-two functions have a natural interpretation as decay functions,
$\Delta^{[\gamma^-]}$ being the probability of a quark to produce a hadron 
in a specific spin state (characterized by the spin vector $S_h$).
Note that the decay probability for an unpolarized quark with nonzero
transverse momentum, can lead to a transverse polarization in the
production of spin 1/2 particles. This transverse polarization is
orthogonal to the quark momentum. Such an effect violates time reversal
invariance. For an out-state this is induced by nonvanishing interactions
in the final state. Indeed, as we will see below, the function 
$\hF_{1T}^\perp$ is a purely interaction-dependent function. 
It has no equivalent in the distribution functions.
The correlation functions $\Delta^{[\gamma^-\gamma_5]}$ and 
$\Delta^{[i\sigma^{i-}\gamma_5]}$ are differences of probabilities for
quarks with different chiralities or transverse spins, respectively, to
produce a hadron in a specific spin state. For the latter also a new decay
function appears because of the non-applicability of time reversal
invariance. A transversely polarized quark with nonzero transverse momentum
can produce unpolarized hadrons, in particular it can produce spinless
particles, such as pions. The relevant function
$H_1^\perp$ is the one appearing in the single spin Collins
effect \cite{Collins-93b}. Note that neutral pions cannot be produced, since
CP-invariance of the strong interactions enforces T-invariance.
 
A useful fragmentation function is the one for a quark (with
momentum $k$)  producing a quark with momentum $p$ ($p^- = z\,k^-$), for
which one has $\Delta^{[\gamma^-]}$ =
$\frac{1}{2}\,\delta(1-z)\,\delta^2(\bpt-\bkt)$. This result is useful to
find cross sections for production of a current jet  with (measured)
transverse momentum $\bpt$ in lepto-production.

The twist analysis for the projections of the quark-quark-gluon matrix 
elements,
\be
\Delta_D^{\alpha [\Gamma ]}(z,\bkt^\prime) =
\left. \int \frac{d\xi^+ d^2\xi_\perp}{4z\,(2\pi)^3} \ e^{i\,k\cdot \xi}\,
Tr\,\langle 0 \vert {\cal L}(0,\xi;n_+)\,\psi(\xi)
\, iD^\alpha (\xi)\,a_h^\dagger a_h\, \overline \psi(0) \Gamma \vert 0 
\rangle \right|_{\xi^- \,=\, 0} ,
\ee
is analogous to the distribution part. 
The correlation function $\Delta_D^{-[\Gamma]}$ is because of the choice of
the path in the link operator trivially related to $k^-\,\Delta^{[\Gamma]}$.
Using the equations of motion one obtains the relations
\begin{eqnarray}
& &
g_{{\scriptscriptstyle T}\alpha\beta}
\,\Delta_{D}^{\alpha [\sigma^{\beta -} ]} =
-\epsilon_{{\scriptscriptstyle T}\alpha \beta}
\,\Delta_D^{\alpha [i\sigma^{\beta -}\gamma_5]}
= i\left(\frac{M_h}{z}\hE - m\,\hF_1 -i\,\frac{M_h}{z}\,\hH
+ i\,\frac{k_T^2} {M_h}\,\hH_1^\perp\right)
\nonumber \\ & & \qquad \qquad \qquad \qquad \qquad \qquad \qquad\qquad
- \epsilon_{{\scriptscriptstyle T}\,ij}\,k_T^i S_{hT}^j \,
(\frac{1}{z}\hH_T^\perp +i\,\frac{m}{M_h}\,\hF_{1T}^\perp),
\\ & &
g_{{\scriptscriptstyle T}\alpha\beta}
\,\Delta_{D}^{\alpha [i\sigma^{\beta -}\gamma_5 ]}
=  \frac{M_h}{z} \hH_s - m\,\hG_{1s}
+ i\,\frac{M_h}{z}\hE_s,
\\ & &
g_{\scriptscriptstyle T}^{\alpha\beta}
\Delta_{D\,\beta}^{[\gamma^{-}]} 
+ i\epsilon_{\scriptscriptstyle T}^{\alpha \beta}\,
\Delta_{D\,\beta}^{\ [\gamma^{-}\gamma_5]}
= k_T^\alpha \left(\frac{1}{z}\hF^\perp +i\,\frac{m}{M_h}
\,\hH_1^\perp\right)
+ i\epsilon_{\scriptscriptstyle T}^{\alpha \beta}k_{T \beta}
\left( \frac{1}{z}\hG_s^\perp - \frac{m}{M_h}\,\hH_{1s}^\perp \right)
\nonumber \\ & & \qquad \qquad \qquad \qquad \qquad \qquad \quad
+ i\epsilon_{\scriptscriptstyle T}^{\alpha \beta}S_{hT\beta} 
\left( \frac{M_h}{z}\hG_T^\prime - m\, \hH_{1T} \right)
\end{eqnarray}
Again it is useful to consider the quantity
\be
\Delta_A^{\alpha [\Gamma]}(z,\bkt^\prime) =
\Delta_D^{\alpha [\Gamma]}(z,\bkt^\prime) 
- k^\alpha\,\Delta^{[\Gamma]}(z,\bkt^\prime),
\ee
which vanishes for $\alpha$ = $-$ and which for transverse indices reduces
after gauge-fixing to the quark-quark-gluon correlation function
(Fig. 3) needed in the calculation of the leptoproduction amplitude.
As for the distribution functions, it is used to define {\em
interaction-dependent} parts and to express the twist-three 
functions into twist-two fragmentation functions and interaction-dependent
parts, explicitly given in Appendix C.

A consistent definition of antiquark fragmentation functions implies
\be
\overline \Delta_{ij}(P_h,S_h;k) = - \Delta_{ij}(P_h,S_h;-k),
\ee
with the symmetry relations $\overline D_1(z,\bkt^{\prime 2})$
= $\hF_1(-z,\bkt^{\prime 2})$ and identically for $\hG_{1T}$, 
$\hH_{1T}$,  $\hH_{1T}^\perp$, $\hE_T$, $\hG_L^\perp$, $\hH$ and $\hH_L$,
while $\overline G_{1L}(z,\bkt^{\prime 2})$
= $-\hG_{1L}(-z,\bkt^{\prime 2})$ and identically for $\hF_{1T}^\perp$, 
$\hH_{1L}^\perp$, $\hH_1^\perp$, $\hE$, $\hF^\perp$, $\hE_L$, 
$\hG_T^\prime$,  $\hG_T^\perp$, $\hH_T^\perp$ and $\hH_T$.

\section{The complete tree-level result}

Up to ${\cal O}(1/Q)$ one needs to include the contributions of
the handbag diagram (Fig. 4), now calculated up to this order with 
in addition
irreducible diagrams with one gluon coupling either to the soft part
involving hadron $H$ or the soft part involving hadron $h$ (see Fig. 5). 
The expressions thus involve the quark-quark-gluon correlation functions.
The momentum-conserving delta-function at the photon vertex is written
(neglecting $1/Q^2$ contributions) as
\be
\delta^4(p+q-k) = \delta(p^++q^+)\,\delta(q^--k^-)
\,\delta^2(\bpt + \bqt - \bkt) ,
\ee
fixing $p^+ = -q^+ = \xbj P^+$ and $k^- = q^- = P_h^-/z_h$. 
The full result, neglecting $1/Q^2$ contributions and still omitting
flavor indices, is then given by
\ba
2M\,{\cal W}_{\mu \nu} & = &
e^2\int dp^-\,dk^+\,d^2\bpt\,d^2\bkt \delta^2(\bpt + \bqt - \bkt) 
\,\Biggl\{
\mbox{Tr} \left( \Phi(p) \gamma_\mu \Delta(k) \gamma_\nu \right)
\nonumber \\
&&\quad \qquad - \mbox{Tr} \left( \gamma_\alpha \frac{{\not\! n}_+}{Q\sqrt{2}}
\gamma_\nu \Phi_A^\alpha (p) \gamma_\mu \Delta (k) \right)
- \mbox{Tr} \left( \gamma_\mu \frac{{\not\! n}_+}{Q\sqrt{2}}
\gamma_\alpha \Delta(k) \gamma_\nu \Phi_A^{\alpha\dagger} (p) \right)
\nonumber \\
&&\quad \qquad
- \mbox{Tr} \left( \gamma_\nu \frac{{\not\! n}_-}{Q\sqrt{2}}
\gamma_\alpha \Phi(p) \gamma_\mu \Delta_A^{\alpha\dagger} (k) \right)
- \mbox{Tr} \left( \gamma_\alpha \frac{{\not\! n}_-}{Q\sqrt{2}}
\gamma_\mu \Delta_A^\alpha (k) \gamma_\nu \Phi (p) \right) \Biggr\}.
\ea
In this expression
the terms with ${\not\! n}_\pm$ arise from fermion propagators
in the hard part neglecting contributions that will appear suppressed
by powers of $Q^2$, i.e.,
\ba
&& \frac{{\not\! p}_1 + {\not\! q} + m}{(p_1 + q)^2} \approx
\frac{(p_1^+ + q^+)\gamma^-}{2(p_1^+ + q^+)q^-} = \frac{\gamma^-}{2q^-}
= \frac{{\not\! n}_+}{Q\sqrt{2}} \approx \frac{\xbj\,{\not\! P}}{Q^2}, \\
&& \frac{{\not\! k}_1 - {\not\! q} + m}{(k_1 - q)^2} \approx
\frac{(k_1^- - q^-)\gamma^+}{-2(k_1^- - q^-)q^+} = \frac{\gamma^+}{(-2q^+)}
= \frac{{\not\! n}_-}{Q\sqrt{2}} \approx \frac{{\not\! P}_h}{z_h \,Q^2}.
\ea
Furthermore, the only relevant quark-quark-gluon correlation functions
are those in which the argument of the gluon field is equal to that of one
of the quark fields, discussed in the previous section. 
These correlation functions can be related to the 
quark-quark correlation functions using the QCD equations of motion. Such
relations are essential to ensure that at order $1/Q$ the sum of diagrams
in Figs 4 and 5 yields an electromagnetically gauge-invariant result. 

The result can be expressed in terms of the twist-two and twist-three
distribution and fragmentation functions and perpendicular tensors and 
vectors. These vectors, $k_\perp^\mu$, $p_\perp^\mu$, $S_\perp^\mu$ and 
$S_{h\perp}^\mu$, are obtained from the expansion into the Cartesian 
directions of $k_T^\mu$, $p_T^\mu$, $S_T^\mu$ and $S_{hT}^\mu$, respectively. 
For that one can use the result
\be
g_T^{\mu\nu} = g_\perp^{\mu \nu}
- \frac{Q_T}{Q} \,\hat q^{\{\mu} \hat x^{\nu\}} 
+ \frac{Q_T}{Q} \,\hat t^{\,\{\mu } \hat x^{\nu\}} ,
\ee
obtained from Eqs~\ref{transverse1} and \ref{transverse2}. The full
expression for the symmetric and antisymmetric parts of the hadronic tensor are,
\begin{eqnarray}
2M\,{\cal W}_S^{\mu\nu} & = &
2z_h\int d^2\bkt\, d^2\bpt\;
\delta^2(\bpt +\bqt-\bkt)\nonumber\\ & &
\mbox{}\ \ \times \Biggl\{
-g_\perp^{\mu \nu} \Biggl[
f_1 \hF_1
+ g_{1s}\hG_{1s}
+ \frac{\epsilon_\perp^{\rho \sigma} k_{\perp\rho} S_{h\perp\sigma}}{M_h}
\,f_1 \hF_{1T}^\perp
\Biggr] \nonumber \\ & & \quad
- \frac{k_\perp^{\{\mu}p_\perp^{\nu\}}
+ (\bk_\perp\cdot \bp_\perp)\,g_\perp^{\mu\nu}}{M M_h}
\,h_{1s}^\perp \hH_{1s}^\perp
- \frac{k_\perp^{\{\mu}S_\perp^{\nu\}}
+ (\bk_\perp\cdot \bS_\perp)\,g_\perp^{\mu\nu}}{M_h}
\,h_{1T} \hH_{1s}^\perp
\nonumber \\ & & \quad
-  \frac{p_\perp^{\{\mu}S_{h\perp}^{\nu\}}
+ (\bp_\perp\cdot \bS_{h\perp})\,g_\perp^{\mu\nu}}{M}
\,h_{1s}^\perp \hH_{1T}
- \left\lgroup S_\perp^{\{\mu}S_{h\perp}^{\nu\}}
+ (\bS_\perp\cdot \bS_{h\perp})\, g_\perp^{\mu\nu} \right\rgroup
\,h_{1T} \hH_{1T}
\nonumber \\ & & \quad
- \frac{\left( k_\perp^{\{\mu} \epsilon_\perp^{\nu \} \rho} p_{\perp\rho}
+  p_\perp^{\{\mu} \epsilon_\perp^{\nu \} \rho} k_{\perp\rho}\right)}{2 M
M_h} \,h_{1s}^\perp \hH_1^\perp
- \frac{\left( k_\perp^{\{\mu} \epsilon_\perp^{\nu \} \rho} S_{\perp\rho}
+  S_\perp^{\{\mu} \epsilon_\perp^{\nu \} \rho} k_{\perp\rho}\right)}{2M_h}
\,h_{1T} \hH_1^\perp
\nonumber \\ & & \quad
+ \frac{2\,\hat t^{\,\{\mu }k_\perp^{\nu\}}}{Q} \Biggl[
 -f_1 \hF_1 + f_1 \frac{\hF^\perp}{z_h}
- g_{1s} \hG_{1s} + g_{1s} \frac{\hG_s^\perp}{z_h}
\nonumber \\ & & \quad \qquad \qquad
+\frac{M}{M_h}\,\xbj \,h_s \hH_{1s}^\perp
-\frac{m}{M_h}\,g_{1s} \hH_{1s}^\perp
- \frac{\bp_\perp\cdot \bS_{h\perp}}{M} \,h_{1s}^\perp \hH_{1T}
\nonumber \\ & & \quad \qquad \qquad
+ \frac{\bp_\perp\cdot \bS_{h\perp}}{M} 
\,h_{1s}^\perp \frac{\hH_T^\perp}{z_h}
- \bS_\perp\cdot \bS_{h\perp} \,h_{1T} \hH_{1T}
+ \bS_\perp\cdot \bS_{h\perp} \,h_{1T} \frac{\hH_T^\perp}{z_h}
\Biggr] \nonumber \\ & & \quad
+ \frac{2\,\hat t^{\,\{\mu }p_\perp^{\nu\}}}{Q} \Biggl[
  \xbj \,f^\perp \hF_1 + \xbj \,g_s^\perp \hG_{1s}
+\frac{M_h}{M}\,h_{1s}^\perp \frac{\hH_s}{z_h}
-\frac{m}{M}\,h_{1s}^\perp \hG_{1s}
+ \frac{\bk_\perp^2}{M M_h}\,h_{1s}^\perp \hH_{1s}^\perp
\nonumber \\ & & \quad \qquad \qquad
+ \frac{\bk_\perp\cdot \bS_{h\perp}}{M} \,h_{1s}^\perp \hH_{1T}
+ \frac{\bk_\perp\cdot \bS_\perp}{M_h}\,\xbj \,h_T^\perp \hH_{1s}^\perp
+ \bS_\perp\cdot \bS_{h\perp} \,\xbj \,h_T^\perp \hH_{1T}
\Biggr] \nonumber \\ & & \quad
+ \frac{2M \,\hat t^{\,\{\mu }S_\perp^{\nu\}}}{Q} \Biggl[
  \xbj \,g_T^\prime \hG_{1s}
+\frac{M_h}{M}\,h_{1T} \frac{\hH_s}{z_h}
-\frac{m}{M}\,h_{1T} \hG_{1s}
+ \frac{\bk_\perp^2}{M M_h}\,h_{1T} \hH_{1s}^\perp
\nonumber \\ & & \quad \qquad \qquad
- \frac{\bk_\perp\cdot \bp_\perp}{M M_h} 
\,\xbj\,h_T^\perp \hH_{1s}^\perp
+ \frac{\bk_\perp\cdot \bS_{h\perp}}{M} \,h_{1T} \hH_{1T}
- \frac{\bp_\perp\cdot \bS_{h\perp}}{M} \,\xbj\,h_T^\perp \hH_{1T}
\Biggr] \nonumber \\ & & \quad
+ \frac{2M_h\,\hat t^{\,\{\mu }S_{h\perp}^{\nu\}}}{Q} \Biggl[
\frac{M}{M_h}\,\xbj \,h_s \hH_{1T}
+ g_{1s} \frac{\hG_T^\prime}{z_h}
-\frac{m}{M_h}\,g_{1s} \hH_{1T}
+ \frac{\bk_\perp\cdot \bp_\perp}{M M_h} \,h_{1s}^\perp \hH_{1T}
\nonumber \\ & & \quad \qquad \qquad
- \frac{\bk_\perp\cdot \bp_\perp}{M M_h} 
\,h_{1s}^\perp \frac{\hH_T^\perp}{z_h}
+ \frac{\bk_\perp\cdot \bS_\perp}{M_h} \,h_{1T} \hH_{1T}
- \frac{\bk_\perp\cdot \bS_\perp}{M_h} \,h_{1T} \frac{\hH_T^\perp}{z_h}
\Biggr] \nonumber \\ & & \quad
+ \frac{2\,\hat t^{\,\{\mu }\epsilon_\perp^{\nu\}\rho}k_{\perp \rho}}{Q} \Biggl[
\frac{M}{M_h}\,\xbj \,h_s \hH_1^\perp
- \frac{m}{M_h}\,g_{1s} \hH_1^\perp
- \frac{\bk_\perp \cdot \bS_{h\perp}}{M_h}\,f_1 \hF_{1T}^\perp
+ \frac{\bp_\perp \cdot \bS_{h\perp}}{M_h}\,\xbj \,f^\perp \hF_{1T}^\perp
\Biggr] \nonumber \\ & & \quad
+ \frac{2\,\hat t^{\,\{\mu }\epsilon_\perp^{\nu\}\rho} p_{\perp\rho}}{Q} \Biggl[
\frac{M_h}{M}\,h_{1s}^\perp \frac{\hH}{z_h}
+ \frac{\bk_\perp^2}{M M_h}\,h_{1s}^\perp \hH_1^\perp
+ \frac{\bk_\perp \cdot \bS_\perp}{M_h}\,\xbj \,h_T^\perp \hH_1^\perp
\Biggr] \nonumber \\ & & \quad
+ \frac{2M\,\hat t^{\,\{\mu}\epsilon_\perp^{\nu\}\rho} S_{\perp\rho}}{Q}\Biggl[ 
\frac{M_h}{M}\,h_{1T} \frac{\hH}{z_h}
+ \frac{\bk_\perp^2}{M M_h}\,h_{1T} \hH_1^\perp
- \frac{\bk_\perp \cdot \bp_\perp}{M M_h}\,\xbj\,h_T^\perp \hH_1^\perp
\Biggr] \nonumber \\ & & \quad
+ \frac{2M_h\,\hat t^{\,\{\mu}\epsilon_\perp^{\nu\}\rho}S_{h\perp\rho}}{Q}
\Biggl[
\frac{\bk_\perp^2}{M_h^2}\,f_1 \hF_{1T}^\perp
- \frac{\bk_\perp \cdot \bp_\perp}{M_h^2}\,\xbj\,f^\perp \hF_{1T}^\perp
\Biggr] \Biggr\}
\label{resulta}
\ea
and
\ba 
2M\,{\cal W}_A^{\mu\nu} & = &
2z_h\int d^2\bkt\, d^2\bpt\;
\delta^2(\bpt +\bqt-\bkt)\nonumber\\ & &
\mbox{}\ \ \times \Biggl\{
i\,\epsilon_\perp^{\mu \nu}\,\Biggl[ f_1 \hG_{1s} + g_{1s} \hF_1 \Biggr]
+ i\frac{k_\perp^{\,[ \mu}S_{h\perp}^{\nu ]}}{M_h}\,g_{1s} \hF_{1T}^\perp
\nonumber \\ & & \quad
+ i\frac{2\,\hat t^{\,[ \mu}k_\perp^{\nu ]}}{Q} \Biggl[
-\frac{M}{M_h}\,\xbj\,e \hH_1^\perp
+\frac{m}{M_h}\,f_1 \hH_1^\perp
+\frac{\bk_\perp \cdot \bS_{h\perp}}{M_h}\,g_{1s} \hF_{1T}^\perp
\nonumber \\ & & \quad \qquad \qquad
+\frac{\bp_\perp \cdot \bS_{h\perp}}{M_h}\,\frac{m}{M}
\,h_{1s}^\perp \hF_{1T}^\perp
-\frac{\bp_\perp \cdot \bS_{h\perp}}{M_h}\,\xbj\,g_s^\perp \hF_{1T}^\perp
\nonumber \\ & & \quad \qquad \qquad
- \bS_\perp \cdot \bS_{h\perp}\,\frac{M}{M_h}\,\xbj\,g_T^\prime \hF_{1T}^\perp
+ \bS_\perp \cdot \bS_{h\perp}\,\frac{m}{M_h}\,h_{1T} \hF_{1T}^\perp
\Biggr] \nonumber \\ & & \quad
+ i\frac{2\,\hat t^{\,[ \mu}p_\perp^{\nu ]}}{Q}\,
\frac{M_h}{M}\,h_{1s}^\perp \frac{\hE_s}{z_h}
+ i\frac{2M\,\hat t^{\,[ \mu}S_\perp^{\nu ]}}{Q}
\,\frac{M_h}{M}\,h_{1T} \frac{\hE_s}{z_h}
\nonumber \\ & & \quad
+ i\frac{2M_h\,\hat t^{\,[ \mu}S_{h\perp}^{\nu ]}}{Q}\Biggl[
-\frac{\bk_\perp^2}{M_h^2}\,g_{1s} \hF_{1T}^\perp
-\frac{\bk_\perp\cdot \bp_\perp}{M M_h}\,\frac{m}{M_h}
\,h_{1s}^\perp \hF_{1T}^\perp
+\frac{\bk_\perp\cdot \bp_\perp}{M_h^2}\,\xbj\,g_s^\perp \hF_{1T}^\perp
\nonumber \\ & & \quad \qquad \qquad
+\frac{\bk_\perp\cdot \bS_\perp}{M_h}\,\frac{M}{M_h}
\,\xbj \,g_T^\prime \hF_{1T}^\perp
-\frac{\bk_\perp\cdot \bS_\perp}{M_h}\,\frac{m}{M_h}\,h_{1T} \hF_{1T}^\perp
\Biggr] \nonumber \\ & & \quad
+ i\frac{2\, \hat t^{\,[\mu}\epsilon_\perp^{\nu]\rho}k_{\perp\rho}}{Q}\Biggl[ 
f_1 \frac{\hG_s^\perp}{z_h}
-f_1 \hG_{1s}
+ g_{1s} \frac{\hF^\perp}{z_h}
- g_{1s} \hF_1
+\frac{M}{M_h}\,\xbj\,e \hH_{1s}^\perp
- \frac{m}{M_h}\,f_1 \hH_{1s}^\perp
\Biggr] \nonumber \\ & & \quad
+ i\frac{2\,\hat t^{\,[\mu}\epsilon_\perp^{\nu]\rho} p_{\perp\rho}}{Q}\Biggl[ 
\xbj\,f^\perp \hG_{1s}
+ \xbj\,g_s^\perp \hF_1
+\frac{M_h}{M}\,h_{1s}^\perp \frac{\hE}{z_h}
- \frac{m}{M}\,h_{1s}^\perp \hF_1
\Biggr] \nonumber \\ & & \quad
+ i\frac{2M\,\hat t^{\,[\mu}\epsilon_\perp^{\nu]\rho}S_{\perp\rho}}{Q}\Biggl[
\xbj \,g_T^\prime \hF_1
+\frac{M_h}{M}\,h_{1T} \frac{\hE}{z_h}
-\frac{m}{M}\,h_{1T} \hF_1
\Biggr] \nonumber \\ & & \quad
+i\frac{2M_h\,\hat t^{\,[\mu}\epsilon_\perp^{\nu]\rho}S_{h\perp\rho}}{Q}\Biggl[
f_1 \frac{\hG_T^\prime}{z_h}
+\frac{M}{M_h}\,\xbj \,e \hH_{1T}
- \frac{m}{M_h}\,f_1 \hH_{1T}
\Biggr] \Biggr\},
\label{resultb}
\ea
where $\{\mu\,\nu\}$ indicates symmetrization of indices and $[\mu\,\nu]$ 
indicates antisymmetrization. Note that one can make the replacement
$\hat t^{\,[\mu}\epsilon_\perp^{\nu]\rho}a_{\perp\rho}$ =
$\epsilon^{\mu\nu\rho\sigma}a_{\perp\rho}\hat q_\sigma$.
The quark distribution functions in hadron $H$ depend on $\xbj $ and $\bpt^2$, 
while the quark fragmentation functions into hadron $h$ depend on $z_h $ and
$\vert \bP_{h\perp}-z\bpt\vert^2$, the squared perpendicular momentum
of hadron $h$ with respect to the quark.

The cross sections are obtained from the
hadronic tensor after contraction with the lepton tensor (Eq.~\ref{contract}).
These contractions, given in Table~\ref{contractions}
use azimuthal angles in the perpendicular plane (see section 2) defined
with respect to the (lepton) scattering plane and the (spacelike) virtual
photon momentum,
\ba
&& \hat x \cdot a_\perp \ = \ - \hat x \cdot \baa_\perp \ = \ -\baa_x \ =
\ - \vert \baa_\perp \vert\, \cos\phi_a, \\
&& \epsilon_\perp^{\mu \nu} \hat x_\mu a_{\perp \nu}
\ \equiv \ \hat x \wedge \baa_\perp \
= \ \baa_y = \vert \baa_\perp \vert\, \sin\phi_a,
\ea
where we have used $\baa_\perp \wedge \bbb_\perp$
$\equiv$ $\epsilon_\perp^{\rho \sigma} a_{\perp\rho} b_{\perp\sigma}$.

\section{Results for leptoproduction integrated over transverse momenta}

After integration over the transverse momenta ($\bP_{h\perp}$ = $-z_h\bqt$) 
the integrations over $\bkt$ and $\bpt$ in Eqs~\ref{resulta} and 
\ref{resultb} can be performed
leading to 
\begin{eqnarray}
2M\,\int d^2P_{h\perp}\,{\cal W}^{\mu\nu} & = &
2z_h\,\Biggl\{ -g_\perp^{\mu \nu} \Bigl[
\,f_1 \hF_1
+ \lambda\lambda_h\,g_1\hG_1\Bigr]
- \left\lgroup S_\perp^{\{\mu}S_{h\perp}^{\nu\}}
+ (\bS_\perp\cdot \bS_{h\perp})\,g_\perp^{\mu\nu}\right\rgroup
h_1 \hH_1
\nonumber \\ & & \mbox{}
+i\,\epsilon_\perp^{\mu \nu}\,\Bigl[ f_1 \lambda_h\,\hG_1
+ \lambda\,g_1 \hF_1 \Bigr] 
+ \lambda_h\,\frac{2M \,\hat t^{\,\{\mu }S_\perp^{\nu\}}}{Q} \Biggl[
  \xbj\,g_T \hG_1 +\frac{M_h}{M}\,h_1 \,\frac{\htH_L}{z_h} \Biggr] 
\nonumber \\ & & \mbox{}
+ \lambda\,\frac{2M_h\,\hat t^{\,\{\mu }S_{h\perp}^{\nu\}}}{Q} \Biggl[
\frac{M}{M_h}\,\xbj\,h_L \hH_1 + g_1 \,\frac{\htG_T}{z_h} \Biggr] 
+ \frac{2M_h\,\hat t^{\,\{\mu} \epsilon_\perp^{\nu\}\rho} S_{\perp \rho}}{Q}
\,h_1 \frac{\htH}{z_h}
\nonumber \\ & & \mbox{}
+ \frac{2M_h\,\hat t^{\,\{\mu} \epsilon_\perp^{\nu\}\rho} S_{h\perp \rho}}{Q}
\,f_1 \,\hF_{1T}^{\perp (1)} 
+ i\lambda_h\,\frac{2M\hat t^{\,[ \mu}S_\perp^{\nu ]}}{Q}\,
\frac{M_h}{M}\,h_1 \frac{\htE_L}{z_h}
\nonumber \\ & & \mbox{}
- i\lambda\,\frac{2M_h\,\hat t^{\,[ \mu}S_{h\perp}^{\nu ]}}{Q}\,
g_1\, \hF_{1T}^{\perp (1)}
+i\,\frac{2M_h\,\hat t^{\,[\mu}\epsilon_\perp^{\nu]\rho}S_{h\perp\rho}}{Q}
\Biggl[ \frac{M}{M_h}\,\xbj \,e \hH_1 + f_1 \,\frac{\htG_T}{z_h} \Biggr] 
\nonumber \\ & & \mbox{}
+i\,\frac{2M\,\hat t^{\,[\mu}\epsilon_\perp^{\nu]\rho}S_{\perp\rho}}{Q}\Biggl[
\xbj \,g_T \hF_1 +\frac{M_h}{M}\,h_1 \,\frac{\htE}{z_h} \Biggr] \Biggr\} ,
\end{eqnarray}
where the quark distribution functions in hadron $H$ depend on $\xbj$, while
the quark fragmentation functions into hadron $h$ depend on $z_h$.
The functions indicated with a tilde appear in the quark-quark-gluon
correlation functions $\Phi_A$ and $\Delta_A$ discussed in section 3. 
These interaction-dependent correlation functions are explicitly
given in Appendix C, just as their $\bkt^\prime$-integrated results.
The function $\hF_{1T}^{\perp (1)}(z)$ is 
the $(\bkt^2/2M_h^2)$-weighted and $\bkt^\prime$-integrated 
result of $\hF_{1T}^\perp(z,\bkt^\prime)$ (see also Appendix C).

Separating the cross sections into parts for unpolarized (O) and 
longitudinally polarized (L) leptons and unpolarized (O), longitudinally
polarized (L) or transversely polarized (T) hadrons in the initial state
one obtains the lepto-production cross section 
\ba
\frac{d\sigma (\vec \ell \vec H \rightarrow \ell^\prime \vec h X)}
{d\xbj\,dy\,dz_h} & = &
\frac{d\sigma_{OO}}{d\xbj\,dy\,dz_h} 
+\frac{d\sigma_{OL}}{d\xbj\,dy\,dz_h} 
+\frac{d\sigma_{OT}}{d\xbj\,dy\,dz_h} 
\nonumber \\ & & \mbox{}
+\frac{d\sigma_{LO}}{d\xbj\,dy\,dz_h} 
+\frac{d\sigma_{LL}}{d\xbj\,dy\,dz_h} 
+\frac{d\sigma_{LT}}{d\xbj\,dy\,dz_h} 
\ea
with (including again the flavor indices)
{
\ba
\frac{d\sigma_{OO}}{d\xbj\,dy\,dz_h} & = &
\frac{4\pi \alpha^2\,s}{Q^4}\,\sum_{a,\bar a} e_a^2 \,\Biggl\{
         \left\lgroup \frac{y^2}{2}+1-y\right\rgroup \xbj f^a_1(\xbj)\,
        \hF^a_1(z_h)
\nonumber \\ && \qquad \qquad 
         + 2\vert \bS_{h\perp}\vert\,(2-y)\sqrt{1-y}\,
          \sin (\phi_s^h)\,\,\frac{M_h}{Q}\,
             \xbj\, f^a_1(\xbj) \hF_{1T}^{\perp (1)\,a}(z_h)
\Biggr\},
\\ 
\frac{d\sigma_{OL}}{d\xbj\,dy\,dz_h} & = &
\frac{4\pi \alpha^2\,s}{Q^4}\,\lambda\,\sum_{a,\bar a} e_a^2 \,\Biggl\{
         \lambda_h \left\lgroup \frac{y^2}{2}+1-y\right\rgroup
         \xbj \,g^a_1(\xbj)\,\hG^a_1(z_h)
\nonumber \\ && \qquad 
         - 2\vert \bS_{h\perp}\vert\,(2-y)\sqrt{1-y}\,
          \cos (\phi_s^h)\,\Biggl[
               \frac{M}{Q}\,\xbj^2\, h^a_L(\xbj) \hH^a_1(z_h)
             + \frac{M_h}{Q}\,\xbj\, g^a_1(\xbj) \frac{\htG^a_T(z_h)}{z_h}
\Biggr]
\Biggr\},
\\
\frac{d\sigma_{OT}}{d\xbj\,dy\,dz_h} & = &
\frac{4\pi \alpha^2\,s}{Q^4}\,\vert \bS_\perp \vert\,\sum_{a,\bar a} e_a^2 
\,\Biggl\{
         2\,(2-y)\sqrt{1-y}\,
          \sin (\phi_s)\,\frac{M_h}{Q}\,\xbj \,h^a_1(\xbj)\,
         \frac{\htH^a(z_h)}{z_h}
\nonumber \\ && \qquad \qquad 
         - 2\lambda_h\,(2-y)\sqrt{1-y}\,
          \cos (\phi_s)\,\Biggl[
               \frac{M}{Q}\,\xbj^2\, g^a_T(\xbj) \hG^a_1(z_h)
             + \frac{M_h}{Q}\,\xbj\, h^a_1(\xbj) \frac{\htH^a_L(z_h)}{z_h}
\Biggr]
\nonumber \\ && \qquad \qquad 
         - \vert \bS_{h\perp} \vert\,(1-y)\,
         \cos (\phi_s + \phi_s^h)\,\,\xbj\,h^a_1(\xbj) \hH^a_1(z_h)
\Biggr\},
\\
\frac{d\sigma_{LO}}{d\xbj\,dy\,dz_h} & = &
\frac{4\pi \alpha^2\,s}{Q^4}\,\lambda_e\,\sum_{a,\bar a} e_a^2 \,\Biggl\{
         \lambda_h \,y\left\lgroup 1-\frac{y}{2} \right\rgroup
         \xbj \,f^a_1(\xbj)\,\hG^a_1(z_h)
\nonumber \\ && \qquad \qquad 
         - 2\,\vert \bS_{h\perp}\vert\,y\sqrt{1-y}\,
          \cos (\phi_s^h)\,\Biggl[
               \frac{M}{Q}\,\xbj^2\, e^a(\xbj) \hH^a_1(z_h)
             + \frac{M_h}{Q}\,\xbj\, f^a_1(\xbj) \frac{\htG^a_T(z_h)}{z_h}
\Biggr]
\Biggr\},
\\
\frac{d\sigma_{LL}}{d\xbj\,dy\,dz_h} & = &
\frac{4\pi \alpha^2\,s}{Q^4}\,\lambda_e\,\lambda\,\sum_{a,\bar a} e_a^2 
\,\Biggl\{
         y \left\lgroup 1-\frac{y}{2} \right\rgroup
             \xbj g^a_1(\xbj)\, \hF^a_1(z_h)
\nonumber \\ && \qquad \qquad \qquad
         + 2\,\vert \bS_{h\perp}\vert\,y\sqrt{1-y}\,
          \sin (\phi_s^h)\,\,\frac{M_h}{Q}\,
             \xbj\, g^a_1(\xbj) \hF_{1T}^{\perp (1)\,a}(z_h)
\Biggr\},
\\
\frac{d\sigma_{LT}}{d\xbj\,dy\,dz_h} & = &
\frac{4\pi \alpha^2\,s}{Q^4}\,\lambda_e\,\vert \bS_\perp \vert
\,\sum_{a,\bar a} e_a^2 \,\Biggl\{
         - 2\,y\sqrt{1-y}\, \cos (\phi_s)\,\Biggl[
               \frac{M}{Q}\,\xbj^2\, g^a_T(\xbj) \hF^a_1(z_h)
             + \frac{M_h}{Q}\,\xbj\, h^a_1(\xbj) \frac{\htE^a(z_h)}{z_h}
\Biggr]
\nonumber \\ && \qquad \qquad \qquad
         - 2\,\lambda_h\,y\sqrt{1-y}\,
          \sin (\phi_s)\,\frac{M_h}{Q}\,\xbj \,h^a_1(\xbj)\,
         \frac{\htE^a_L(z_h)}{z_h}
\Biggr\}.
\ea }
The familiar inclusive cross section is easily obtained by using the
results for the fragmentation function of a quark into a quark and summing
over the quark spins,
\be
\frac{d\sigma (\vec \ell \vec H \rightarrow \ell^\prime X)}
{d\xbj\,dy} =
\frac{d\sigma_{OO}}{d\xbj\,dy} 
+\frac{d\sigma_{LL}}{d\xbj\,dy} 
+\frac{d\sigma_{LT}}{d\xbj\,dy} 
\ee
with
\ba
\frac{d\sigma_{OO}}{d\xbj\,dy} & = &
\frac{4\pi \alpha^2\,s}{Q^4}\,\sum_{a,\bar a} e_a^2 \,
         \left\lgroup \frac{y^2}{2}+1-y\right\rgroup \xbj f^a_1(\xbj)
\\ 
\frac{d\sigma_{LL}}{d\xbj\,dy} & = &
\frac{4\pi \alpha^2\,s}{Q^4}\,\lambda_e\,\lambda\,\sum_{a,\bar a} e_a^2 
         \,y \left\lgroup 1-\frac{y}{2} \right\rgroup
             \xbj g^a_1(\xbj)
\\
\frac{d\sigma_{LT}}{d\xbj\,dy} & = &
-\frac{4\pi \alpha^2\,s}{Q^4}\,\lambda_e\,\vert \bS_\perp \vert
\,\sum_{a,\bar a} e_a^2 \,
         2\,y\sqrt{1-y}\, \cos (\phi_s)\,
               \frac{M}{Q}\,\xbj^2\, g^a_T(\xbj). 
\ea
These inclusive results lead to the tree-level results for the structure
functions $F_1(\xbj,Q^2)$, $F_2(\xbj,Q^2)$, 
$\mbox{\Large {\em g}}_1(\xbj,Q^2)$ and $\mbox{\Large {\em g}}_2(\xbj,Q^2)$
which were given in the Introduction. Comparing semi-inclusive and
inclusive results one finds the familiar result for the number of produced
particles 
\be
N_h(\xbj,z_h) =
\frac{\frac{d\sigma_{OO}}{d\xbj\,dy\,dz_h}}{\frac{d\sigma_{OO}}{d\xbj\,dy}}
=\frac{\sum_{a,\bar a} e_a^2\,f_1^a(\xbj)\,\hF_1^a(z_h)}{\sum_{a,\bar a} e_a^2\,
f_1^a(\xbj)}.
\ee
Asymmetries in the number of produced particles for polarized leptons
and/or polarized target hadrons are obtained from the cross sections by
putting $S_h$ to zero, 
{
\be
N_h(\xbj,z_h;\lambda_e;\lambda,\bS_\perp) =
N_h(\xbj,z_h)\,\left[ 1 + \vert \bS_\perp \vert \sin(\phi_s)\,\Delta
N_{h,OT} + \lambda_e\lambda\,\Delta N_{h,LL} + \lambda_e\vert \bS_\perp
\vert\,\cos(\phi_s)\,\Delta N_{h,LT}\right],
\ee }
involving the nonvanishing ones of the following asymmetries,
{
\ba
\Delta N_{h,OL}(\xbj,z_h) & = &
\frac{\sigma(\ell \stackrel{\rightarrow}{H}) 
- \sigma(\ell \stackrel{\leftarrow}{H})}
{\sigma(\ell \stackrel{\rightarrow}{H}) 
+ \sigma(\ell \stackrel{\leftarrow}{H})}
= \frac{\sigma_{OL}}{\sigma_{OO}} = 0, 
\label{eq95}\\
\Delta N_{h,OT}(\xbj,z_h) & = &
\frac{\sigma(\ell \stackrel{\rightarrow}{H}) 
- \sigma(\ell \stackrel{\leftarrow}{H})}
{\sigma(\ell \stackrel{\rightarrow}{H}) 
+ \sigma(\ell \stackrel{\leftarrow}{H})}
= \frac{\sigma_{OT}}{\sigma_{OO}}  
= \frac{2(2-y)\sqrt{1-y}}{1-y+\frac{1}{2}y^2}\,\frac{M_h}{Q}
\,\frac{\sum_{a,\bar a} e_a^2\,h_1^a(\xbj)\,\htH^a(z_h)/z_h}
{\sum_{a,\bar a} e_a^2\,f_1^a(\xbj)\,\hF_1^a(z_h)}, 
\label{eq96}\\
\Delta N_{h,LO}(\xbj,z_h) & = &
\frac{\sigma(\stackrel{\rightarrow}{\ell} H) 
- \sigma(\stackrel{\leftarrow}{\ell} H)}
{\sigma(\stackrel{\rightarrow}{\ell} H) 
+ \sigma(\stackrel{\leftarrow}{\ell} H)}
= \frac{\sigma_{LO}}{\sigma_{OO}} = 0, 
\label{eq97}\\
\Delta N_{h,LL}(\xbj,z_h) & = &
\frac{\sigma(\stackrel{\rightarrow}{\ell} \stackrel{\rightarrow}{H}) 
- \sigma(\stackrel{\leftarrow}{\ell} \stackrel{\rightarrow}{H}) 
- \sigma(\stackrel{\rightarrow}{\ell} \stackrel{\leftarrow}{H}) 
+\sigma(\stackrel{\leftarrow}{\ell} \stackrel{\leftarrow}{H})}
{\sigma(\stackrel{\rightarrow}{\ell} \stackrel{\rightarrow}{H}) 
+ \sigma(\stackrel{\leftarrow}{\ell} \stackrel{\rightarrow}{H}) 
+ \sigma(\stackrel{\rightarrow}{\ell} \stackrel{\leftarrow}{H}) 
+\sigma(\stackrel{\leftarrow}{\ell} \stackrel{\leftarrow}{H})}
= \frac{\sigma_{LL}}{\sigma_{OO}}
= \frac{y(1-\frac{1}{2}y)}{1-y+\frac{1}{2}y^2}\,
\frac{\sum_{a,\bar a} e_a^2\,g_1^a(\xbj)\,\hF_1^a(z_h)}
{\sum_{a,\bar a} e_a^2\,f_1^a(\xbj)\,\hF_1^a(z_h)},
\label{eq98}\\
\Delta N_{h,LT}(\xbj,z_h)  & = &
\frac{\sigma(\stackrel{\rightarrow}{\ell} \stackrel{\rightarrow}{H}) 
- \sigma(\stackrel{\leftarrow}{\ell} \stackrel{\rightarrow}{H}) 
- \sigma(\stackrel{\rightarrow}{\ell} \stackrel{\leftarrow}{H}) 
+\sigma(\stackrel{\leftarrow}{\ell} \stackrel{\leftarrow}{H})}
{\sigma(\stackrel{\rightarrow}{\ell} \stackrel{\rightarrow}{H}) 
+ \sigma(\stackrel{\leftarrow}{\ell} \stackrel{\rightarrow}{H}) 
+ \sigma(\stackrel{\rightarrow}{\ell} \stackrel{\leftarrow}{H}) 
+\sigma(\stackrel{\leftarrow}{\ell} \stackrel{\leftarrow}{H})}
= \frac{\sigma_{LT}}{\sigma_{OO}}
\nonumber \\
& = & \frac{2(2-y)\sqrt{1-y}}{1-y+\frac{1}{2}y^2}\,\left[
\frac{M}{Q}\,\frac{\sum_{a,\bar a} e_a^2\,\xbj\,g_T^a(\xbj)\,\hF_1^a(z_h)}
{\sum_{a,\bar a} e_a^2\,f_1^a(\xbj)\,\hF_1^a(z_h)}
+\frac{M_h}{Q}\,\frac{\sum_{a,\bar a} e_a^2\,
\xbj\,h_1^a(\xbj)\,\htE^a(z_h)/z_h} {\sum_{a,\bar a} e_a^2\,
f_1^a(\xbj)\,\hF_1^a(z_h)}\right]. 
\label{eq99}
\ea }
For the double asymmetries, $\Delta N_{h,LL}$ and
$\Delta N_{h,LT}$ it is (at least in principle) sufficient to
consider either the asymmetry for the electron spins or for the target spins 
since $\Delta N_{h,OL} = \Delta N_{h,LO} = 0$. Although $\Delta N_{h,OT} \ne 0$,
the asymmetries in $\Delta N_{h,OT}$ and $\Delta N_{h,LT}$ involve different
transverse directions.
Of the asymmetries in the number of produced hadrons only $\Delta N_{h,LL}$ 
is leading. The other asymmetries are
proportional to $1/Q$. For a target with transverse polarization in the
scattering plane the result is proportional to the transverse spin
distribution $h_1^a$ and the twist-three fragmentation function $\htH^a$.
For a target with transverse polarization orthogonal to the scattering
plane the result contains the product of the twist-three distribution
$g_T^a$ with the ordinary fragmentation function $\hF_1^a$ and the product of
the transverse spin distribution $h_1^a$ with the twist-three fragmentation
function $\htE^a$ \cite{Jaffe-Ji-93}. We note that in the naive parton 
model approach the tilde functions do not appear at tree level.

Measuring polarization in the final state, gives several new possibilities
to study quark-quark and quark-quark-gluon correlation functions. As
discussed in Appendix A the coefficient of $S_h$ in $d\sigma/d\xbj dy dz_h$
determines the polarization. It is convenient to parametrize the
polarization vector ${\cal P}_h^\mu$ in the same way as $S_h^\mu$
(Eq.~\ref{spinh}) with lightcone helicity $\Lambda_h$ and transverse
polarization $\bbox{\cal P}_{h\perp}$. The polarization induced in the
final state is
{
\ba
&&\Lambda_h\,N_h(\xbj,z_h;\lambda_e;\lambda,\bS_\perp)
= \lambda\,\frac{\sum_{a,\bar a} e_a^2
\,g_1^a(\xbj)\,\hG_1^a(z_h)}{\sum_{a,\bar a} e_a^2\,f_1^a(\xbj)}
+ \lambda_e\,\frac{y(1-\frac{1}{2}y)}{1-y+\frac{1}{2}y^2}
\,\frac{\sum_{a,\bar a} e_a^2\,
f_1^a(\xbj)\,\hG_1^a(z_h)}{\sum_{a,\bar a} e_a^2\,f_1^a(\xbj)}
\nonumber \\ && \qquad \qquad \qquad \mbox{}
- \bS_x\,\,\frac{2(2-y)\sqrt{1-y}}{1-y+\frac{1}{2}y^2}
\,\left[ \frac{M}{Q}
\,\frac{\sum_{a,\bar a} e_a^2\,\xbj\, g^a_T(\xbj) \hG^a_1(z_h)}
{\sum_{a,\bar a} e_a^2\,f_1^a(\xbj)}+\frac{M_h}{Q}
\,\frac{\sum_{a,\bar a} e_a^2\,h^a_1(\xbj) \htH^a_L(z_h)/z_h}
{\sum_{a,\bar a} e_a^2\,f_1^a(\xbj)} \right]
\nonumber \\ && \qquad \qquad \qquad \mbox{}
- \bS_y\,\,\frac{2y\sqrt{1-y}}{1-y+\frac{1}{2}y^2}
\,\frac{M_h}{Q}\,
\frac{\sum_{a,\bar a} e_a^2\,h^a_1(\xbj)\,\htE^a_L(z_h)/z_h}
{\sum_{a,\bar a} e_a^2\,f_1^a(\xbj)},
\label{eq100}
\\
&&\bbox{\cal P}_{hx}\,N_h(\xbj,z_h;\lambda_e;\lambda,\bS_\perp) =
-\bS_x\,\frac{1-y}{1-y+\frac{1}{2}y^2}\,
\frac{\sum_{a,\bar a} e_a^2
\,h_1^a(\xbj)\,\hH_1^a(z_h)}{\sum_{a,\bar a} e_a^2\,f_1^a(\xbj)}
\nonumber \\ && \qquad \qquad \qquad \mbox{}
- \lambda\,\,\frac{2(2-y)\sqrt{1-y}}{1-y+\frac{1}{2}y^2}
\,\left[ \frac{M}{Q}
\,\frac{\sum_{a,\bar a} e_a^2\,\xbj\, h^a_L(\xbj) \hH^a_1(z_h)}
{\sum_{a,\bar a} e_a^2\,f_1^a(\xbj)} + \frac{M_h}{Q}
\,\frac{\sum_{a,\bar a} e_a^2\, g^a_1(\xbj) \htG^a_T(z_h)/z_h}
{\sum_{a,\bar a} e_a^2\,f_1^a(\xbj)} \right]
\nonumber \\ && \qquad \qquad \qquad \mbox{}
- \lambda_e\,\,\frac{2y\sqrt{1-y}}{1-y+\frac{1}{2}y^2}
\,\left[ \frac{M}{Q}
\,\frac{\sum_{a,\bar a} e_a^2\,\xbj\,e^a(\xbj) \hH^a_1(z_h)}
{\sum_{a,\bar a} e_a^2\,f_1^a(\xbj)} + \frac{M_h}{Q}
\,\frac{\sum_{a,\bar a} e_a^2\,f^a_1(\xbj)\htG^a_T(z_h)/z_h}
{\sum_{a,\bar a} e_a^2\,f_1^a(\xbj)} \right],
\label{eq101}
\\
&&\bbox{\cal P}_{hy}\,N_h(\xbj,z_h;\lambda_e;\lambda,\bS_\perp) =
\bS_y\,\frac{1-y}{1-y+\frac{1}{2}y^2}\,
\frac{\sum_{a,\bar a} e_a^2
\,h_1^a(\xbj)\,\hH_1^a(z_h)}{\sum_{a,\bar a} e_a^2\,f_1^a(\xbj)}
\nonumber \\ && \qquad \qquad \qquad \qquad \qquad \mbox{}
+ \frac{2(2-y)\sqrt{1-y}}{1-y+\frac{1}{2}y^2}\,
\frac{M_h}{Q}\,
\frac{\sum_{a,\bar a} e_a^2\,
f_1^a(\xbj)\,\hF_{1T}^{\perp (1)\,a}(z_h)}{\sum_{a,\bar a} e_a^2\,
f_1^a(\xbj)}
\nonumber \\ && \qquad \qquad \qquad \qquad \qquad \mbox{}
+\lambda_e\,\lambda\,\,
\frac{2y\sqrt{1-y}}{1-y+\frac{1}{2}y^2}\,
\frac{M_h}{Q}\,
\frac{\sum_{a,\bar a} e_a^2\,
g_1^a(\xbj)\,\hF_{1T}^{\perp (1)\,a}(z_h)}{\sum_{a,\bar a} e_a^2\,f_1^a(\xbj)}.
\label{eq102}
\ea }
In leading order the final state polarization is proportional to the 
corresponding polarization (longitudinal or transverse) of the target
hadron and the polarization (longitudinal) of the lepton. The transfer
parameters involve the leading distribution functions $f_1^a$, $g_1^a$ and
$h_1^a$ and the leading fragmentation functions $\hG_1^a$ and $\hH_1^a$.
For instance, the measurement of transversely polarized $\Lambda$'s has been
proposed \cite{Artru-91} as a way to obtain the transverse spin 
distribution $h_1^a$.

In the complete result up to order $1/Q$ final state polarization arises
from several other polarizations in the initial state. Especially
noteworthy is the (transverse) polarization in the production of a spin 1/2
particle (e.g.\ a $\Lambda$) for an unpolarized initial state, the term
involving $f_1^a(\xbj)\,\hF_{1T}^{\perp (1)\,a}(z_h)$ in Eq.~\ref{eq102}.
This polarization is purely transverse, orthogonal to the lepton scattering
plane. It involves the (interaction-dependent) time-reversal odd
fragmentation function $\hF_{1T}^{\perp\,a}$.

At order $1/Q$ one finds also transverse polarization in the production of
hadrons, coming from longitudinal polarization in the initial state or
vice versa. This offers new possibilities to obtain twist-three distribution
functions such as $g_T^a(x)$ and also $h_L^a(x)$. For the latter
this semi-inclusive polarization transfer measurement is the equivalent
of a double-spin asymmetry measurement in Drell-Yan 
scattering \cite{Jaffe-Ji-93,Tangerman-Mulders-94}. We note the presence
of several interaction-dependent functions which complicate the naive parton
picture for these observables.

\section{Azimuthal asymmetries}

The transverse momentum dependence in Eqs~\ref{resulta} and \ref{resultb} lead 
after integration over $\bkt$ and $\bpt$ to a dependence on
$\bqt = -\bP_{h\perp}/z$. Furthermore transverse directions appear in
the spin vectors. It is useful to project the momenta $\bp_\perp$ and 
$\bk_\perp$ onto the direction $\bhh$,
\ba
&&
a_\perp^\mu \ = \
(\bhh\cdot \baa_\perp)\,\hat h^\mu
+ (\bhh \wedge \baa_\perp)\, \epsilon_\perp^{\mu \rho}\,\hat h_\rho, \\
&&
\epsilon_\perp^{\mu \rho}\,a_{\perp \rho} \ =
\ -(\bhh \wedge \baa_\perp)\,\hat h^\mu +
(\bhh\cdot \baa_\perp)\, \epsilon_\perp^{\mu \rho}\,\hat h_\rho.
\ea
We will not consider polarization in the final state.
The result for the symmetric and antisymmetric parts in the hadronic tensor
are, 
{
\ba 
2M\,{\cal W}_S^{\mu\nu} & = & 2z_h\,\Biggl\{
-g_\perp^{\mu \nu}\, I[f_1 \hF_1] 
\nonumber \\ & & 
+ \frac{2\hat t^{\,\{\mu }\hat h_\perp^{\nu\}}}{Q} \Biggl(
 I\left[(\bhh\cdot \bk_\perp)\,f_1 
\left\lgroup \frac{\hF^\perp}{z_h }-\hF_1 \right\rgroup \right] 
+ \xbj\,I\left[(\bhh\cdot \bp_\perp)\,f^\perp \hF_1\right]
\Biggr)
\nonumber \\ & & 
-\lambda\,\Biggl\{
\hat h^{\,\{\mu} \epsilon_\perp^{\nu\}\rho} \hat h_\rho\,
\,I\left[\,\frac{2\,(\bhh\cdot \bk_\perp)(\bhh\cdot \bp_\perp)
-\bk_\perp\cdot\bp_\perp}{M M_h} \,\, h_{1L}^\perp \hH_1^\perp \right]
\nonumber \\ & & \qquad
- 2\hat t^{\,\{\mu} \epsilon_\perp^{\nu \} \rho} \hat h_\rho
\,\Biggl( \frac{M}{Q} \,I\Bigl[ \frac{\bhh\cdot \bk_\perp}{M_h}
\left\lgroup \xbj \,h_L - \frac{m}{M}\,g_{1L}\right\rgroup
 \hH_1^\perp \Bigr]
 \nonumber \\ & & \qquad\qquad\qquad\quad
 + \frac{M_h}{Q}\, I\Bigl[ \frac{\bhh\cdot \bp_\perp}{M}\,h_{1L}^\perp
\left\lgroup \frac{\hH}{z_h } + \frac{\bk_\perp^2}{M_h^2}
\,\hH_1^\perp\right\rgroup
\Bigr] \Biggr) \Biggr\}
\nonumber \\ & & 
-\frac{1}{2} \left( \hat h^{\,\{\mu}\epsilon_\perp^{\nu\}\rho} S_{\perp\rho}
+ S_\perp^{\,\{\mu}\epsilon_\perp^{\nu\}\rho} \hat h_\rho \right)
\, I\Bigr[ \frac{(\bhh\cdot \bk_\perp)}{M_h}\,\,
h_1 \hH_1^\perp \Bigr]
\nonumber \\ & & \mbox{} 
-\hat h^{\,\{\mu}\epsilon_\perp^{\nu\}\rho}\hat h_\rho
\,(\bhh\cdot\bS_\perp)\,
I\Bigr[\frac{4\,(\bhh\cdot \bp_\perp)^2 (\bhh\cdot \bk_\perp)
-\bp_\perp^2\,(\bhh\cdot \bk_\perp)
-2\,(\bhh\cdot \bp_\perp)(\bk_\perp\cdot \bp_\perp)}{2\,M^2\,M_h}\,\,
h_{1T}^\perp \hH_1^\perp \Bigr]
\nonumber \\ & & \mbox{} 
+ \left\lgroup 2\,\hat h^\mu \hat h^\nu + g_\perp^{\mu \nu} \right\rgroup
(\bhh\wedge\bS_\perp)\, 
\nonumber \\ & & \qquad \qquad \mbox{}
\times\,I\Bigr[ \frac{2\,(\bhh\cdot \bp_\perp)(\bk_\perp\cdot \bp_\perp)
+ \bp_\perp^2\,(\bhh\cdot \bk_\perp)
-4\,(\bhh\cdot \bp_\perp)^2 (\bhh\cdot \bk_\perp)}{2\,M^2\,M_h}\,\,
h_{1T}^\perp \hH_1^\perp \Bigr]
\nonumber \\ & & 
\mbox{} + 2\hat t^{\,\{\mu} \epsilon_\perp^{\nu\} \rho} S_{\perp \rho} 
\Biggl( \frac{M}{Q}\,
I\,\Bigl[\frac{\bk_\perp \cdot \bp_\perp}{2M M_h}\,
\left\lgroup \xbj h_T - \frac{m}{M}\,g_{1T} - \xbj h_T^\perp
\right\rgroup \hH_1^\perp \Bigr]
\nonumber \\ & & \qquad \qquad \qquad \qquad \quad \mbox{}
+\frac{M_h}{Q}\,I\,\Bigl[ h_1 \left\lgroup \frac{\hH}{z_h }
+ \frac{\bk_\perp^2}{M_h^2}\,\hH_1^\perp \right\rgroup \Bigr]
\Biggr)
\nonumber \\ & & 
\mbox{} + 2\hat t^{\,\{\mu}\epsilon_\perp^{\nu\}\rho} \hat h_\rho
\,(\bhh\cdot \bS_\perp)\,\Biggl(
\frac{M_h}{Q}\,I\Bigl[
\frac{2\,(\bhh\cdot \bp_\perp)^2-\bp_\perp^2}{2M^2}\,
h_{1T}^\perp\, \left\lgroup \frac{\hH}{z_h } 
+ \frac{\bk_\perp^2}{M_h^2}\,\hH_1^\perp \right\rgroup
\Bigr]
\nonumber \\ & & \qquad \qquad \qquad \mbox{}
+ \frac{M}{Q}\,I\Bigl[
\frac{2\,(\bhh\cdot \bk_\perp)(\bhh\cdot \bp_\perp)
- \bk_\perp\cdot \bp_\perp}{2MM_h}
\left\lgroup \xbj h_T - \frac{m}{M}\,g_{1T} + \xbj h_T^\perp
\right\rgroup \hH_1^\perp \Bigr]
\Biggr)
\nonumber \\ & &
\mbox{} + 2\hat t^{\,\{\mu} \hat h^{\nu\}}
\,(\bhh\wedge \bS_\perp)\,\Biggl(
\frac{M_h}{Q}\,I\Bigl[
\frac{2\,(\bhh\cdot \bp_\perp)^2 -\bp_\perp^2}{2M^2} 
\,h_{1T}^\perp\, \left\lgroup \frac{\hH}{z_h } 
+ \frac{\bk_\perp^2}{M_h^2}\,\hH_1^\perp \right\rgroup
\Bigr]
\nonumber \\ & & \qquad \qquad \qquad \mbox{}
- \frac{M}{Q}\,I\Bigl[
\frac{\bk_\perp\cdot \bp_\perp
- 2\,(\bhh\cdot\bk_\perp)(\bhh\cdot \bp_\perp)}{2MM_h}
\left\lgroup \xbj h_T - \frac{m}{M}\,g_{1T} + \xbj h_T^\perp
\right\rgroup \hH_1^\perp \Bigr]
\Biggr) \Biggr\}
\ea }
and
{
\ba
2M\,{\cal W}_A^{\mu\nu} & = & 2\,z_h\,\Biggl\{
i\,\lambda\,\epsilon_\perp^{\mu \nu}\, I\Bigl[ g_{1L} \hF_1 \Bigr]
+ i\,\epsilon_\perp^{\mu \nu}\,(\bhh\cdot \bS_\perp )\,
I\Bigl[ \frac{\bhh\cdot \bp_\perp}{M}\,g_{1T} \hF_1 \Bigr]
\nonumber \\ &&
-2i\,\hat t^{\,[ \mu}\hat h^{\nu ]}\,\frac{M}{Q}
\,I\Bigl[ \frac{\bhh\cdot \bk_\perp}{M_h}
\left\lgroup \xbj \,e -\frac{m}{M}\,f_1 \right\rgroup  \hH_1^\perp \Bigr]
\nonumber \\ &&
+ 2i\,\lambda\,\hat t^{\,[\mu} \epsilon_\perp^{\nu]\rho}\hat h_\rho\,\Biggl(
\frac{M_h}{Q}\,I\Bigl[ \frac{\bhh\cdot \bk_\perp}{M_h}\,g_{1L}
\left \lgroup \frac{\hF^\perp}{z_h } - \hF_1 \right\rgroup \Bigr]
+\frac{M}{Q}\,I\Bigl[ \frac{\bhh\cdot \bp_\perp}{M}\,\xbj g_L^\perp
\hF_1\Bigr]
\nonumber \\ & & \qquad\qquad\qquad\qquad\qquad\mbox{}
+\frac{M_h}{Q}\, I\Bigl[ \frac{\bhh\cdot \bp_\perp}{M}\,h_{1L}^\perp
\left\lgroup \frac{\hE}{z_h } - \frac{m}{M} \hF_1 \right\rgroup \Bigr]
\Biggr) 
\nonumber \\ & & \mbox{}
+ 2i\,\hat t^{\,[\mu}\epsilon_\perp^{\nu]\rho} \hat h_\rho 
\,(\bhh\cdot \bS_\perp )\,\Biggl(
\frac{M_h}{Q}\,I\Bigl[
\frac{2(\bhh\cdot\bp_\perp)(\bhh\cdot\bk_\perp)-\bp_\perp\cdot
\bk_\perp}{2MM_h}\,g_{1T}
\left\lgroup \frac{\hF^\perp}{z_h } - \hF_1 \right\rgroup \Bigr]
\nonumber \\ & & \qquad\qquad\qquad\qquad\qquad\mbox{}
+ \frac{M}{Q}\,I\Bigl[
\frac{2(\bhh\cdot\bp_\perp)^2-\bp_\perp^2}{2M^2}\,\xbj  g_T^\perp \hF_1 \Bigr]
\nonumber \\ & & \qquad\qquad\qquad\qquad\qquad\mbox{}
+\frac{M_h}{Q}\,I\Bigl[
\frac{2(\bhh\cdot\bp_\perp)^2-\bp_\perp^2}{2M^2} h_{1T}^\perp
\left\lgroup \frac{\hE}{z_h }- \frac{m}{M}\, \hF_1 \right\rgroup \Bigr]\Biggr)
\nonumber \\ & & \mbox{}
+ 2i\,\hat t^{\,[\mu} \hat h^{\nu]}\,(\bhh\wedge \bS_\perp )\,\Biggl(
\frac{M_h}{Q}\,I\Bigl[
\frac{2(\bhh\cdot\bp_\perp)(\bhh\cdot\bk_\perp)-\bp_\perp\cdot
\bk_\perp}{2MM_h}\,g_{1T}
\left\lgroup \frac{\hF^\perp}{z_h } - \hF_1 \right\rgroup \Bigr]
\nonumber \\ & & \qquad\qquad\qquad\qquad\qquad\mbox{}
+ \frac{M}{Q}\,I\Bigl[
\frac{2(\bhh\cdot\bp_\perp)^2-\bp_\perp^2}{2M^2}\,\xbj  g_T^\perp \hF_1 \Bigr]
\nonumber \\ & & \qquad\qquad\qquad\qquad\qquad\mbox{}
+ \frac{M_h}{Q}\,I\Bigl[
\frac{2(\bhh\cdot\bp_\perp)^2-\bp_\perp^2}{2M^2} h_{1T}^\perp
\left\lgroup \frac{\hE}{z_h }- \frac{m}{M}\, \hF_1 \right\rgroup \Bigr]\Biggr)
\nonumber \\ & & \mbox{}
 + 2i\,\hat t^{\,[\mu} \epsilon_\perp^{\nu]\rho} S_{\perp\rho} \,\Biggl(
\frac{M}{Q}\,I\Bigl[ \xbj  g_T \hF_1 \Bigr]
+\frac{M_h}{Q}\,I\Bigl[ h_1
\left\lgroup \frac{\hE}{z_h }- \frac{m}{M}\, \hF_1 \right\rgroup \Bigr]
\nonumber \\ & & \qquad\qquad\qquad\qquad\qquad\mbox{}
+\frac{M_h}{Q}\,I\Bigl[ \frac{\bp_\perp\cdot \bk_\perp}{2MM_h}\, g_{1T}
\left\lgroup \frac{\hF^\perp}{z_h }- \hF_1 \right\rgroup \Bigr]
\Biggr)\Biggr\} ,
\ea }
with (including again flavor indices) convolution integrals of the type
\ba
I\left[(\bhh\cdot \bp_\perp)\,f\, \hF\right] (\xbj,z_h,Q_T)
& = & \sum_{a,\bar a} e_a^2\int d^2\bpt\, 
(\bhh\cdot \bpt)\,f^a(\xbj,\bpt^2) 
\hF^a(z_h,\vert \bP_{h\perp} - z_h\bpt\vert^2).
\ea
For a gaussian transverse momentum dependence the relevant integrals are
evaluated in Appendix D. In principle one can now identify structure 
functions by looking at the most general expression for the hadronic tensor
\cite{Meng-Olness-Soper-92}. We will omit this step. 
Rather, we immediately give the cross section following from the above
tensor. We will use the results for gaussian transverse momentum dependence.
The complete result
for the cross section is again written as a sum over terms
depending on lepton (O and L) and target hadron (O, L and T) polarizations.
The contributions that are nonzero in leading order in $1/Q$ are
{
\ba
&&\frac{d\sigma_{OO}}{d\xbj\,dy\,dz_h\,d^2\bP_{h\perp}} \ = \ 
\frac{4\pi \alpha^2\,s}{Q^4}\,\sum_{a,\bar a} e_a^2\,
\left\lgroup \frac{y^2}{2}+1-y\right\rgroup  \,\xbj f_1^a(\xbj)\,\hF^a_1(z_h)
\,\frac{{\cal G}(Q_T;R)}{z_h^2},
\label{soo}
\\
&&\frac{d\sigma_{OL}}{d\xbj\,dy\,dz_h\,d^2\bP_{h\perp}} \ = \ 
-\frac{4\pi \alpha^2\,s}{Q^4}\,\lambda\,\sum_{a,\bar a} e_a^2\,
(1-y)\, \sin (2\phi_h)\,\frac{Q_T^2\,R^4}{M M_h\,R_H^2\,R_h^2}
\,\xbj h_{1L}^{\perp\,a}(\xbj) \hH_1^{\perp\,a}(z_h) 
\,\frac{{\cal G}(Q_T;R)}{z_h^2},
\\
&&\frac{d\sigma_{OT}}{d\xbj\,dy\,dz_h\,d^2\bP_{h\perp}} \ = \ 
-\frac{4\pi \alpha^2\,s}{Q^4}\,\vert \bS_\perp\vert
\,\sum_{a,\bar a} e_a^2\,\Biggl\{
(1-y)\,\sin(\phi_h + \phi_s) \,\frac{Q_T\,R^2}{M_h\,R_h^2}
\,\xbj h^a_1(\xbj) \hH_1^{\perp\,a}(z_h)
\nonumber \\ && \qquad\qquad\qquad\qquad\qquad\qquad
+(1-y) \,\sin(3\phi_h - \phi_s) \,
\frac{Q_T^3\,R^6}{2M^2M_h\,R_H^4\,R_h^2}
\,\xbj h_{1T}^{\perp\,a}(\xbj) \hH_1^{\perp\,a}(z_h) 
\Biggr\}\,\frac{{\cal G}(Q_T;R)}{z_h^2},
\label{sot}
\\
&&\frac{d\sigma_{LL}}{d\xbj\,dy\,dz_h\,d^2\bP_{h\perp}} \ = \ 
\frac{4\pi \alpha^2\,s}{Q^4}\,\lambda_e\lambda\,\sum_{a,\bar a} e_a^2\,
 y\left(1-\frac{y}{2}\right) \,\xbj\,g^a_{1L}(\xbj)\,\hF^a_1(z_h) 
\,\frac{{\cal G}(Q_T;R)}{z_h^2},
\\
&&\frac{d\sigma_{LT}}{d\xbj\,dy\,dz_h\,d^2\bP_{h\perp}} \ = \ 
\frac{4\pi \alpha^2\,s}{Q^4}\,\lambda_e\,\vert \bS_\perp\vert
\,\sum_{a,\bar a} e_a^2\,
y\left(1-\frac{y}{2}\right)\,\cos (\phi_h - \phi_s)\,\frac{Q_T\,R^2}{M\,R_H^2}
\,\xbj\,g^a_{1T}(\xbj) \hF^a_1(z_h) 
\,\frac{{\cal G}(Q_T;R)}{z_h^2},
\ea }
where ${\cal G}(Q_T;R)$ = $(R^2/\pi)\,\exp(-Q_T^2R^2)$, i.e., a gaussian
of which the fall-off is determined by a radius $R$. This radius is
related to the radii $R_H$ and $R_h$ governing the fall-off of $f(x,\bkt^2)$
and $D(z,\bkt^{\prime 2})$ as $R^2$ = $R_H^2 R_h^2/(R_H^2 + R_h^2)$.
These radii again may depend on the longitudinal momentum and on the
specific function, i.e., $R_H$ = $R_H^f(\xbj)$ and $R_h$ = $R_h^D(z_h)$.
Note that in leading order $\sigma_{LO}$ = 0.
These results have been discussed in Refs \cite{Kotzinian-95} and 
\cite{Tangerman-Mulders-95b}. The possibility to measure $h_1^a$ by 
considering asymmetry in the production of hadrons (e.g.\ pions), the
first term in Eq.~\ref{sot}, was first discussed by Collins \cite{Collins-93b}.
The complete results show that the measurements of specific azimuthal
asymmetries in deep-inelastic lepton-hadron scattering with polarized 
beam and target allow the measurement of all six ($x$- and $k_T$-dependent) 
quark distributions, of course keeping in mind that also a separation of the
different flavors is needed. For obtaining a first indication for the
behavior of the results, the dominance of u-quarks in the proton,
especially at moderate $x$-values will be helpful. In all the expressions
only two fragmentation functions play a role. The first is the 'ordinary'
fragmentation function $\hF_1^a$, the second is the time-reversal odd
function $\hH_1^{\perp a}$.  

In principle the general expression for the hadronic tensor in the beginning
of this section allows the calculation of all azimuthal asymmetries in the
cross section that contribute up to order $1/Q$. Below, we give first the 
results for unpolarized spin 1/2 targets,
which for unpolarized leptons is the extension of Eq.~\ref{soo},
{
\ba
&&\frac{d\sigma_{OO}}{d\xbj\,dy\,dz_h\,d^2\bP_{h\perp}} \ = \ 
\frac{4\pi \alpha^2\,s}{Q^4}\,\sum_{a,\bar a} e_a^2\,\Biggl\{
\left\lgroup \frac{y^2}{2}+1-y\right\rgroup  \,\xbj f_1^a(\xbj)\,\hF^a_1(z_h)
\nonumber \\ && \qquad
- 2(2-y)\sqrt{1-y}\, \cos (\phi_h)\, \frac{Q_T}{Q} \Biggl( 
\,\frac{R^2}{R_H^2}\,\xbj^2 f^{\perp\,a}(\xbj) \hF^a_1(z_h)
-\frac{R^2}{R_h^2}\,\xbj f^a_1(\xbj)\,\frac{\htF^{\perp\,a}(z_h)}{z_h}  
\Biggr)  \Biggr\}\,\frac{{\cal G}(Q_T;R)}{z_h^2},
\\
&&\frac{d\sigma_{LO}}{d\xbj\,dy\,dz_h\,d^2\bP_{h\perp}} \ = \ 
-\frac{4\pi \alpha^2\,s}{Q^4}\,\lambda_e\,\sum_{a,\bar a} e_a^2\,
2y\sqrt{1-y}\, \sin \phi_h\,
\frac{Q_T}{Q}\,\frac{M\,R^2}{M_h\,R_h^2}\,
\xbj^2 \,\tilde e^a(\xbj) \,\hH_1^{\perp\,a}(z_h)
\,\frac{{\cal G}(Q_T;R)}{z_h^2},
\ea }
The $\langle \cos(\phi_h)\rangle$ asymmetry in unpolarized leptoproduction,
unfortunately is rather complicated, involving one twist-three
distribution function ($f^{\perp a}$) and one twist-three fragmentation 
function ($\hF^{\perp a}$) \cite{Levelt-Mulders-94a}. 
It is important to point out, however, that
the $\langle \cos(\phi_h)\rangle$ asymmetry is not only a kinematical
effect. It reduces to a kinematical factor only depending on $y$ and $Q^2$
when the interaction-dependent pieces in the
twist-three functions are set to zero, $\tilde f^{\perp a} = 0$ and
$\htF^{\perp a} = 0$, implying $f^{\perp a} = f_1^a/\xbj$ and 
$\hF^{\perp a} = z_h\,\hF_1^{\perp a}$.
At order $1/Q$ there is no $\langle \cos(2\phi_h)\rangle$ asymmetry in
the deep-inelastic leptoproduction cross section. For polarized leptons
and unpolarized targets a $\langle \sin(\phi_h)\rangle$ asymmetry is
found \cite{Levelt-Mulders-94b}, involving the interaction-dependent
part of the distribution function $e^a$ and the time-reversal odd
fragmentation function $\hH_1^{\perp a}$. Noteworthy is that it is the same 
fragmentation function that appears in several of the leading azimuthal 
asymmetries for polarized targets.

As indicated, the above asymmetries have been discussed in various papers.
The remaining asymmetries at order $1/Q$, i.e.\ those for polarized targets,
can also be extracted from the general result at the beginning of this 
section. They are
{
\ba
&&\frac{d\sigma_{OL}}{d\xbj\,dy\,dz_h\,d^2\bP_{h\perp}} \ = \ 
-\frac{4\pi \alpha^2\,s}{Q^4}\,\lambda\,\sum_{a,\bar a} e_a^2\,\Biggl\{
(1-y)\, \sin (2\phi_h)\,\frac{Q_T^2\,R^4}{M M_h\,R_H^2\,R_h^2}
\,\xbj h_{1L}^{\perp\,a}(\xbj) \hH_1^{\perp\,a}(z_h) 
\nonumber \\ && \qquad\qquad\qquad
- 2 (2-y)\sqrt{1-y}\, \sin (\phi_h)\,\frac{Q_T}{Q}\,\Biggl( 
\frac{R^6}{MM_h\,R_H^4R_h^4}\left\lgroup \frac{R_h^2-R_H^2}{R^2} - Q_T^2R_h^2
\right\rgroup\,\xbj h_{1L}^{\perp\,a}(\xbj)\,\hH_1^{\perp\,a}(z_h)
\nonumber \\ && \qquad\qquad\qquad\qquad\qquad\qquad
+\frac{M\,R^2}{M_h\,R_h^2}\,\xbj^2 \tilde h^a_L(\xbj)\,\hH_1^{\perp\,a}(z_h)
- \frac{M_h\,R^2}{M\,R_H^2}\,\xbj h_{1L}^{\perp\,a}(\xbj) 
\,\frac{\htH^a(z_h)}{z}\Biggr)
\Biggr\}\,\frac{{\cal G}(Q_T;R)}{z_h^2},
\\
&&\frac{d\sigma_{OT}}{d\xbj\,dy\,dz_h\,d^2\bP_{h\perp}} \ = \ 
-\frac{4\pi \alpha^2\,s}{Q^4}\,\vert \bS_\perp\vert
\,\sum_{a,\bar a} e_a^2\,\Biggl\{
(1-y)\,\sin(\phi_h + \phi_s) \,\frac{Q_T\,R^2}{M_h\,R_h^2}
\,\xbj h^a_1(\xbj) \hH_1^{\perp\,a}(z_h)
\nonumber \\ && \qquad\qquad\qquad
+(1-y) \,\sin(3\phi_h - \phi_s) \,
\frac{Q_T^3\,R^6}{2M^2M_h\,R_H^4\,R_h^2}
\,\xbj h_{1T}^{\perp\,a}(\xbj) \hH_1^{\perp\,a}(z_h) 
\nonumber \\ && \qquad\qquad\qquad
- 2(2-y)\sqrt{1-y}\,\sin (\phi_s)\,\Biggl(
\frac{Q_T}{Q}\,\frac{Q_T\,R^8}{2M^2M_h\,R_H^6R_h^4}
\left\lgroup \frac{2R_H^2-R_h^2}{R^2} + Q_T^2R_h^2 \right\rgroup
\,\xbj h_{1T}^{\perp\,a}(\xbj)\,\hH_1^{\perp\,a}(z_h)
\nonumber \\ && \qquad\qquad\qquad \qquad\qquad\qquad\qquad
+\frac{M_h}{Q}\,\xbj h^a_1(\xbj) \,\frac{\htH^a(z_h)}{z_h} 
- \frac{Q_T}{Q}\,\frac{Q_T\,R^4}{2M_h\,R_H^2R_h^2}
\xbj^2\left( \tilde h_T^a(\xbj) + \tilde h_T^{\perp\,a}(\xbj) \right)
\hH_1^{\perp\,a}(z_h) \Biggr)
\nonumber \\ && \qquad\qquad\qquad
+ 2(2-y)\sqrt{1-y}\,\sin (2\phi_h - \phi_s)\, \Biggl(
\frac{Q_T}{Q}\,\frac{Q_T\,R^4}{2M_h\,R_H^2R_h^2}\,
\xbj \,\left(2h_1^a(\xbj)- \xbj \tilde h_{1L}^{\perp\,a}(\xbj)\right) 
\,\hH_1^{\perp\,a}(z_h) 
\nonumber \\ && \hspace{7.5 cm}
-\frac{M_h}{Q}\,\frac{Q_T^2\,R^4}{2M^2\,R_H^4}\,
\xbj h_{1T}^{\perp\,a}(\xbj) \frac{\htH^a(z_h)}{z_h}
\Biggr) \Biggr\}\,\frac{{\cal G}(Q_T;R)}{z_h^2},
\\
&&\frac{d\sigma_{LL}}{d\xbj\,dy\,dz_h\,d^2\bP_{h\perp}} \ = \ 
\frac{4\pi \alpha^2\,s}{Q^4}\,\lambda_e\lambda\,\sum_{a,\bar a} e_a^2\,\Biggl\{
 y\left(1-\frac{y}{2}\right) \,\xbj\,g^a_{1L}(\xbj)\,\hF^a_1(z_h) 
\nonumber \\ && \qquad\qquad\qquad
 -2y\sqrt{1-y}\,\cos(\phi_h)\,
\frac{Q_T}{Q}\,\Biggl(
\frac{R^2}{R_H^2}\,\xbj^2 g_L^{\perp\,a}(\xbj) \hF^a_1(z_h)
-\frac{R^2}{R_h^2}\,\xbj\,g_{1L}^a(\xbj) \,\frac{\htF^{\perp\,a}(z_h)}{z_h}
\nonumber \\ && \hspace{8 cm}
+\frac{M_h\,R^2}{M\,R_H^2}\, \xbj h_{1L}^{\perp\,a}(\xbj)
\frac{\htE^a(z_h)}{z_h} \Biggr)
\Biggr\}\,\frac{{\cal G}(Q_T;R)}{z_h^2},
\\
&&\frac{d\sigma_{LT}}{d\xbj\,dy\,dz_h\,d^2\bP_{h\perp}} \ = \ 
\frac{4\pi \alpha^2\,s}{Q^4}\,\lambda_e\,\vert \bS_\perp\vert
\,\sum_{a,\bar a} e_a^2\,\Biggl\{
y\left(1-\frac{y}{2}\right)\,\cos (\phi_h - \phi_s)\,\frac{Q_T\,R^2}{M\,R_H^2}
\,\xbj\,g^a_{1T}(\xbj) \hF^a_1(z_h) 
\nonumber \\ && \qquad\qquad\qquad
-2y\sqrt{1-y}\,\cos (\phi_s)\,\Biggl(
\frac{M}{Q}\,\xbj^2 g^a_T(\xbj) \hF^a_1(z_h) 
+\frac{M_h}{Q}\,\xbj h^a_1(\xbj) \,\frac{\htE^a(z_h)}{z_h} 
\nonumber \\ & & \hspace{6 cm}
+\frac{M_h}{Q}\,\frac{R^2}{2MM_h\,R_H^2R_h^2}\left\lgroup 1 - Q_T^2\,R^2
\right\rgroup \xbj g^a_{1T}(\xbj)\,\frac{\htF^{\perp\,a}(z_h)}{z_h}
\Biggr)
\nonumber \\ && \qquad\qquad \qquad 
-2y\sqrt{1-y}\,\cos(2\phi_h-\phi_s)\,\frac{Q_T}{Q}\,\Biggl(
\frac{Q_T\,R^4}{2M\,R_H^4}\,\xbj^2 g_T^{\perp\,a}(\xbj)\, \hF^a_1(z_h) 
+\frac{Q_T\,R^4}{2M\,R_H^4}\,\frac{M_h}{M}\,
\xbj h_{1T}^{\perp\,a}(\xbj) \,\frac{\htE^a(z_h)}{z_h} 
\nonumber \\ && \hspace{8 cm}
-\frac{Q_T\,R^4}{2M\,R_H^2R_h^2}\,
\xbj g^a_{1T} \,\frac{\htF^{\perp\,a}(z_h)}{z_h} \Biggr)
\Biggr\}\,\frac{{\cal G}(Q_T;R)}{z_h^2}.
\ea }
Because the results involve mostly higher harmonics in the $\phi_h$-dependence
and involve most of the allowed combinations of twist-two and twist-three
distribution and fragmentation functions discussed in section 3, we do not
expect that these results soon will be used to extract the new information
they contain on the structure of the nucleon, specifically the 
quark-quark-gluon correlations contained in the interaction dependent
pieces. In first instance, the result have been given because they can
be used to estimate azimuthal asymmetries. Such estimates may be necessary
in the absence of full azimuthal coverage. We note that the estimate of
the asymmetries in the naive parton model is obtained by setting all
interaction-dependent pieces (functions with a tilde) to zero.
In that case the twist-three functions can be expressed 
in the twist-two ones. The explicit relations are given in Appendix C. 
Note, however, that although the functions appear in the $1/Q$ parts
of the cross section, there is no reason to expect that the 
interaction-dependent functions themselves are smaller than the 
twist-two functions.  They are just different matrix elements involving 
gluon fields.

Finally, we want to discuss the case in which only the azimuthal direction
of the current jet is detected, without an analysis of this jet. The
measurement of the (small) transverse momentum of the current jet allows (in
principle) a direct study of the transverse momentum of the distribution
functions. The starting point are Eqs~\ref{resulta} and \ref{resultb} in 
which one uses the quark $\rightarrow$ quark fragmentation function.
The nonvanishing contributions in the cross section 
(with $\bp_\perp$ being the jet transverse momentum involving azimuthal 
angle $\phi_j$) are
\begin{eqnarray}
&&\frac{d\sigma_{OO}}{d\xbj\,dy\,d^2\bp_\perp} =
\frac{4\pi \alpha^2\,s}{Q^4}\,\sum_{a,\bar a} e_a^2\,\Biggl\{
\left\lgroup \frac{y^2}{2}+1-y\right\rgroup  \,\xbj f_1^a(\xbj,\bp_\perp^2)
\nonumber \\ && \qquad \qquad \qquad \qquad \qquad \qquad
- 2(2-y)\sqrt{1-y}\, \cos (\phi_j)\,\frac{\vert \bp_\perp\vert}{Q} 
\,\xbj^2 f^{\perp\,a}(\xbj,\bp_\perp^2) \Biggr\},
\\
&&\frac{d\sigma_{LL}}{d\xbj\,dy\,d^2\bp_\perp} \ = \ 
\frac{4\pi \alpha^2\,s}{Q^4}\,\lambda_e\lambda\,\sum_{a,\bar a} e_a^2\,\Biggl\{
 y\left(1-\frac{y}{2}\right) \,\xbj\,g^a_{1L}(\xbj,\bp_\perp^2)
\nonumber \\ && \qquad \qquad \qquad \qquad \qquad \qquad
- 2y\sqrt{1-y}\, \cos (\phi_j)\,\frac{\vert \bp_\perp\vert}{Q} 
\,\xbj^2 g_L^{\perp\,a}(\xbj,\bp_\perp^2) \Biggr\}.
\\
&&\frac{d\sigma_{LT}}{d\xbj\,dy\,d^2\bp_\perp} \ = \ 
\frac{4\pi \alpha^2\,s}{Q^4}\,\lambda_e\vert \bS_\perp\vert\,
\sum_{a,\bar a} e_a^2\,\Biggl\{
 y\left(1-\frac{y}{2}\right) \,\frac{\vert \bp_\perp \vert}{M}\,
\cos(\phi_j-\phi_s)\,\xbj\,g^a_{1T}(\xbj,\bp_\perp^2)
\nonumber \\ && \qquad \qquad \qquad \qquad \qquad \qquad
- 2y\sqrt{1-y}\, \cos (\phi_s)\,\frac{M}{Q} 
\,\xbj^2 g_T^a(\xbj,\bp_\perp^2) 
\nonumber \\ && \qquad \qquad \qquad \qquad \qquad \qquad
- 2y\sqrt{1-y}\, \cos (2\phi_j-\phi_s)
\,\frac{M}{Q}\,\frac{\vert \bp_\perp\vert^2}{2M^2} 
\,\xbj^2 g_T^{\perp\,a}(\xbj,\bp_\perp^2) \Biggr\}.
\end{eqnarray} 
Again decomposing the twist-three functions in these jet asymmetries into their
twist-two parts and the interaction-dependent piece, 
one obtains the naive parton model results, if the interaction-dependent 
pieces are set to zero. For the unpolarized case, the cross section then
becomes proportional to $f_1^a$, reducing the $\langle \cos(\phi_j)\rangle$ 
asymmetry to a purely kinematical effect depending only on $y$ and $Q$
as discussed in Ref. \cite{Cahn-78}. 
Similarly, in the case of longitudinal polarization setting the 
interaction-dependent pieces to zero, produces a result proportional to
the quark helicity distribution $g_1^a$ plus a quark mass term.
Except for this mass term the $\langle \cos(\phi_j)\rangle$ asymmetry
is purely kinematical in that case. We would like to turn things around,
however, and state the importance of finding deviations of
the naive parton results for unpolarized and polarized jet production in 
order to get an idea of the quark-quark-gluon correlation functions 
$\tilde f^{\perp a}$ and $\tilde g_L^{\perp a}$ in a nucleon. 
For transversely polarized hadrons one can study $g_T^a(\xbj,\bp_\perp^2)$,
which integrated over $\bp_\perp$ can also be measured in inclusive
leptoproduction. 

\section{Summary}

In this paper the complete tree-level result up to order $1/Q$ for polarized
deep-inelastic leptoproduction has been presented. The formalism is the
diagrammatic approach in which soft hadronic parts are represented by
expectation values of nonlocal - here bilocal - combinations of quark
and gluon fields, referred to as correlation functions. Both quark-quark
and quark-quark-gluon correlation functions need to be included in order
$1/Q$. The quark-quark-gluon correlation functions can be related to the 
quark-quark correlation functions using the QCD equations of motion. Such
relations are essential to ensure an electromagnetically gauge-invariant 
result. 
Essential in our treatment is also the explicit treatment of quark transverse
momenta in order to study azimuthal asymmetries in semi-inclusive
lepto-production processes.

The full result for the hadronic tensor contains a large number of terms.
They are some of the allowed terms in the most general expansion of the
hadronic tensor, which can be expressed in terms of 'standard' tensor
structures built from the momenta and spin vectors of the hadrons involved
multiplied with structure functions. Having such an expansion one could
read off the result for the structure functions. In view of the large
number of structure functions that are allowed in principle, we found it
more practical to split up the cross section in the parts involving the
lepton polarizations (unpolarized or longitudinal polarization) and
hadron polarizations (unpolarized, longitudinal and transverse polarization).

In the leptoproduction cross section in which one integrates over all
transverse momenta the results have been presented as the numbers of produced
hadrons $N_h(\xbj,z_h)$ and the asymmetries in these numbers arising from the
polarization of initial state lepton or hadron. Finally the polarization
of a produced (spin 1/2) hadron has been discussed. The results include
among others the following features
\begin{itemize}
\item
An asymmetry in produced particles (e.g.\ pions) is obtained at order $1/Q$ for
unpolarized leptons and transversely polarized hadrons (Eq.~\ref{eq96}) 
proportional to the transverse spin distribution $h_1^a$ and a 
twist-three fragmentation function.
\item
A transverse polarization orthogonal to the scattering plane is induced in
spin 1/2 hadrons in the final state (e.g.\ $\Lambda$-baryons) starting with
an unpolarized lepton and unpolarized target hadron. This polarization is
proportional to the unpolarized quark distributions $f_1^a$ and the
fragmentation function $\hF_{1T}^{\perp (1)\,a}$ (Eq.~\ref{eq102}). For
longitudinally polarized lepton and target hadron a similar polarization
emerges, now proportional to the quark helicity distribution $g_1^a$ and
the same fragmentation function $\hF_{1T}^{\perp (1)\,a}$. Although the
fragmentation function is a new one, allowed because of the non-applicability
of time-reversal invariance in the production, the comparison of induced
polarization for polarized and unpolarized initial state may give another
handle on determination of quark helicity distributions, especially
because the production of $\Lambda$'s produces different flavor weighting
compared with inclusive processes.
\item
For completeness also all leading asymmetries in number of particles and in
induced polarization have been
included in Eqs~\ref{eq95}-\ref{eq99} and Eqs~\ref{eq100}-\ref{eq102}.
These are the ones that appear in the naive parton model.
\end{itemize}

The dependence on quark transverse momenta becomes explicit in semi-inclusive
lepto-production in the transverse momentum of the produced hadrons. Here
the following features emerge,
\begin{itemize}
\item
All six twist-two $x$- and $\bpt$-dependent quark distribution functions for
a spin 1/2 hadron can be accessed in leading order asymmetries if one 
considers lepton and hadron polarizations. One of the asymmetries involves
the transverse spin distribution $h_1^a$. On the production side, only two
different fragmentation functions are involved, the familiar unpolarized
fragmentation function $\hF_1^a$ and the (interaction-dependent) fragmentation
function $\hH_1^{\perp a}$.
\item
At order $1/Q$ a $\langle \sin(\phi_h)\rangle$ asymmetry is found for
polarized leptons and unpolarized hadrons. This probes the 
interaction-dependent distribution functions $\tilde e^a$ in combination
with the same fragmentation function $\hH_1^{\perp a}$ mentioned in the
previous item.
\end{itemize}
Several more azimuthal asymmetries are found for polarized targets. The results
have been given explicitly because they indicate the type of asymmetries
to be expected in deep-inelastic leptoproduction. At this point we also
want to point out once more that our results represent tree-level
calculations. The inclusion of gluon ladder-graphs give corrections 
proportional to $\alpha_s$ and $\alpha_s\,\ln Q^2$. 
For the cross sections integrated over
transverse momenta, they lead to scale-dependence of the distribution
functions. They will also enter as additional contributions to several
of the observables, including the ones that are zero at tree-level. A
well-known example of the latter is the longitudinal structure function
$F_L = F_2 - 2\xbj F_1$, which is proportional to $\alpha_s$. Work along
these lines with explicit treatment of transverse momenta of quarks
is under investigation.

We acknowledge contributions of J. Levelt (University of 
N\"urnberg-Erlangen) in the early stage of this work.
We thank D. Boer and R. Jacob for a careful reading of the manuscript.
This work is part of the research program of the foundation for
Fundamental Research of Matter (FOM) and the National Organization
for Scientific Research (NWO).

\appendix

\section{Spin vectors}

In this appendix we would like to remind the reader of the role of the spin 
vectors $S$ and $S_h$ for initial state and final state hadrons, respectively. 
A more extended discussion of aspects related to the role of
the spin vector of the produced hadron (applied in $e^+e^-$ annihilation) can
be found in Ref.~\cite{Chen-et-al-95}.
Starting with the quantities
\begin{eqnarray*}
\tilde {\cal W}^{\mu\nu}_{ij,kl}(q,P,P_h) & = & 
\langle h;k\vert T\vert H;i\rangle\,\langle H;j\vert T^\dagger\vert h;l\rangle
\\& \quad - \quad & 
\mbox{generalized\ tensor\ in\ $2\times 2$\ H-spin-space\ (ij)\ 
and\ in\ $2\times 2$\ h-spin-space\ (kl),}
\nonumber \\
\rho_{ij}(P,S) & = & \langle H;i\vert \sum_\alpha \vert \alpha \rangle
p(\alpha) \langle \alpha \vert H;j\rangle
\quad \stackrel{\mbox{H\ rest\ frame}}{\equiv} \quad \frac{1}{2}\Bigl(
\bbox{1}_{ij} +  \bbox{\sigma}_{ij}\cdot \bS \Bigr)
\\ &  \quad - \quad & \mbox{initial\ state\ density\ matrix\ ,\ defining
\ $S$},
\nonumber \\ 
R^\ast_{kl}(P_h,f) & = & \langle h;k\vert f\rangle\,\langle f\vert h;l\rangle
\quad \stackrel{\mbox{h\ rest\ frame}}{\equiv} \quad w(f)\,\Bigl( \bbox{1}_{kl}
+  \bbox{\sigma}^\ast_{kl}\cdot \bS_h \Bigr)
\\ & - &  \mbox{decay\ matrix\ into\ final\ states\ f
\ in\ $2\times 2$\ h-spin-space\ (kl),\ defining\ $S_h$,}
\end{eqnarray*}
where the latter is normalized to $\sum_f R^\ast_{kl}(f) = \delta_{kl}$.
Note that the summation over $f$ is just symbolic for all kinematic variables
appearing in the decay.
An example of $w(f)$ is the center-of-mass distribution $w(\theta_{\pi N})$ in
the decay of the $\Lambda$ baryon.
Starting with some (general) initial state, one has the following result for 
the semi-inclusive hadronic tensor in which the decay products of hadron $h$
are detected,
\be
{\cal W}^{\mu\nu}(i\rightarrow f) = 
\rho_{ji}(P,S)\,\tilde{\cal W}^{\mu\nu}_{ij,kl}(q,P,P_h)\,R_{lk}(P_h,f)
\equiv \tilde {\cal W}^{\mu\nu}_{kl}(q;P,S;P_h)\,R_{lk}(P_h,f).
\ee
This shows for instance why the spin vector, originating from $\rho(P,S)$ 
can only appear linearly in the
parametrization of the tensor. Summing (or integrating) over the
decay products of $h$ one is not able to measure the polarization of $h$, 
\be
\sum_{f}{\cal W}^{\mu\nu}(i\rightarrow h \rightarrow f)
= \tilde {\cal W}^{\mu\nu}_{kk}(q;P,S;P_h) 
\equiv {\cal W}^{\mu\nu}(q;P,S;P_h,S_h = 0). 
\ee
The general result in the h-spin-space can (in the rest-frame of hadron
h) be written as the unpolarized
result times the production density matrix parametrized in terms of
the polarization $\bbox{\cal P}_h$,
\be
\tilde {\cal W}^{\mu\nu}_{kl}(q;P,S;P_h) 
\equiv {\cal W}^{\mu\nu}(q;P,S;P_h,0)\,\frac{1}{2}\Bigl( \bbox{1}_{kl}
+ \bbox{\sigma}_{kl}\cdot \bbox{\cal P}_h(q;P,S;P_h) \Bigr) .
\ee
Thus one can write in the $h$-rest frame, 
\ba 
{\cal W}^{\mu\nu}(i\rightarrow f) & = &
{\cal W}^{\mu\nu}(q;P,S;P_h,0)\,w(f)\,\Bigl( 1 + \bbox{\cal
P}(q;P,S;P_h)\cdot  \bS_h(P_h;f)\Bigr)
\nonumber \\ & \equiv & {\cal W}^{\mu\nu}(q;P,S;P_h,S_h)\,w(f).
\ea
This shows that also the spin vector $S_h$ appears only linearly, while
it multiplies the polarization vector as appearing in the production matrix.
Parametrizing the polarization vector ${\cal P}_h$
similar as $S_h$ in a longitudinal polarization $\Lambda_h$ and a transverse
polarization $\bbox{\cal P}_{hT}$, which because the polarization vector
parametrizes a density matrix satisfy $\Lambda_h^2 
+ \bbox{\cal P}_{hT}^2 \le 1$,
one can in a general frame write for $\bbox{\cal P}\cdot \bS_h$ =
$\Lambda_h\,\lambda_h + \bbox{\cal P}_{h\perp}\cdot \bS_{h\perp}$.

\section{The link operator}

In this appendix we discuss the path structure of the link operator used
in the definition of the various correlation functions. In hard processes
one encounters the correlation functions integrated over at least one
lightcone component of the parton momentum. The actual component which one
needs is actually fixed by the large vector $q$. This vector can be used
to fix a lightlike vector for each given (small, i.e.\ $p^2 \ll Q^2$) vector
$p$, 
\be
n_p \propto \frac{1}{Q}\left( p + \frac{2\,p\cdot q}{Q^2}\,q\right).
\ee
For the correlation function $\Phi(P,S;p)$ one obtains as a consequence of
the fact that $p^2 \sim P^2 \sim p\cdot P \ll Q^2$
\be
n_p \approx n_P = \left\lgroup \frac{Q}{A},\ 0,\ \bnull_T \right\rgroup
\ = \ n_-.
\ee
The lightlike vector is (approximately, i.e., up to corrections of
${\cal O}(1/Q^2)$) equal for all momenta in a 'connected' soft part. These
momenta themselves define the 'other' lightlike vector, $p/Q \propto P/Q
\propto n_+$.
Similarly one finds for $\Delta(P_h,S_h;k)$ that
\be
n_k \approx n_{P_h} = \left\lgroup 0,\ \frac{A}{Q},\ \bnull_T
\right\rgroup \ = \ n_+, 
\ee
while $k/Q \propto P_h/Q \propto n_-$.
The lightlike vector determines the integration used, for instance in 
$\int dp^-\,\Phi(P,S;p)$ the component $p^- = (Q/A)\,p\cdot n_+$. It also fixes
the path structure in the link operator
\be
{\cal L}(0,x;n_-) = {\cal P} \exp \left(-ig\int_0^{x} ds^\mu 
\,A_\mu(s)\right),
\ee
The link operator ${\cal L}(0,x;n_-)$  is actually 
constructed as the average of two links with paths $P_1$ and $P_2$. 
For $\Phi$ both lie in the $x^+ = 0$ plane and run along the $n_-$
direction except for endpoints at $x^- = \pm \infty$. The path $P_1$
runs from $[0,0,\bnull_T]$ to  
$[x^-,0,\bxt]$ via  $[+\infty, 0, \bnull_T]$ and $[+\infty, 0, \bxt]$,
while the path $P_2$
runs from $[0,0,\bnull_T]$ to  $[x^-,0,\bxt]$ via  $[-\infty, 0, \bnull_T]$
and $[-\infty, 0, \bxt]$. In the
gauge $A^+ = n_-\cdot A$ = 0, the link operator in the projections
$\Phi^{[\Gamma]}$ becomes unity, except for the parts at $x^- = \pm
\infty$. However, it is well-known that $A^+ = 0$ does not completely fix
the gauge. The residual gauge freedom is for instance fixed by imposing an
antisymmetric boundary condition on $A_T$ \cite{Kogut-Soper-70}.
Fixing in this way the residual gauge freedom on $A_T$, 
$A_T[-\infty,x^+,\bxt]$ = $-A_T[\infty,x^+,\bxt]$,
allows a unique inversion of the relation $G^{+i}$ = $\partial^+ A_T^i$, 
\be
A_T^i[x^-,x^+,\bxt] = \frac{1}{2} \int dy^-\,\epsilon(x^- - y^-)\,
G^{+i}[y^-,x^+,\bxt] \ \equiv \ \frac{1}{\partial^+}G^{+i}[x^-,x^+,\bxt].
\ee
The choice ${\cal L}(0,x;n_-)$ = (${\cal L}_1(0,x;P_1)$
+ ${\cal L}_2(0,x;P_2)$)/2 then reduces to unity.

For the link operator in the correlation functions $\Delta$ all choices are 
based on the $n_+$-direction. It becomes simply unity in the gauge fixed
by $A^- = 0$ and an antisymmetric boundary condition for $A_T$.

\section{Explicit quark-quark-gluon correlation functions}

For the correlation function,
\be
\Phi_{A}^{\alpha [\Gamma]}(x,\bpt) 
 =  \Phi_D^{\alpha [\Gamma]}(x,\bpt) - p^\alpha\,\Phi^{[\Gamma]}(x,\bpt),
\ee
one has $\Phi_A^{+ [\Gamma]} = 0$ while the functions with transverse 
indices reduce after gauge fixing ($A^+ = 0$) to a quark-quark-gluon 
matrix element (Eq.~\ref{eq50}).
It is convenient to parametrize the following combinations (in which all
indices $\alpha$ and $\beta$ are transverse only) as
\ba
&&\Phi_{A}^{\alpha [i\sigma^{\beta +}\gamma_5]}-\Phi_{A}^{\beta
[i\sigma^{\alpha +}\gamma_5]}
= i\epsilon_{\scriptscriptstyle T}^{\alpha \beta}\, Mx\,\tilde e 
- \left(S_T^\alpha p_T^\beta - p_T^\alpha S_T^\beta \right)
\,x\,\tilde h_T^\perp,
\\ & &
g_{{\scriptscriptstyle T}\alpha\beta}
\,\Phi_{A}^{\alpha [i\sigma^{\beta +}\gamma_5 ]}
= M x \,\tilde h_s,
\\ & &
\Phi_{A}^{\alpha [\gamma^{+}]}
- i\epsilon_{\scriptscriptstyle T}^{\alpha \beta}\,
\Phi_{A\,\beta}^{\ [\gamma^{+}\gamma_5]}
= p_T^\alpha \,x\,\tilde f^\perp 
- i\epsilon_{\scriptscriptstyle T}^{\alpha \beta}p_{T\beta} 
\,x\tilde g_s^\perp 
- i\epsilon_{\scriptscriptstyle T}^{\alpha \beta}S_{T\beta} 
\,Mx\,\tilde g_T^\prime ,
\ea
in terms of (interaction-dependent) functions indicated with a tilde.
Using the results for $\Phi_D$ as following from the equations of motion
(see section 3), one can split the twist-three distribution functions into the 
twist-two part and an interaction-dependent part. 
The first part is the one which is obtained in the naive parton model. 
The complete list is
\ba
e(x,\bpt^2) & = & \frac{m}{M}\,\frac{f_1(x,\bpt^2)}{x} 
+ \tilde e(x,\bpt^2)  , \\
f^\perp(x,\bpt^2) & = & \frac{f_1(x,\bpt^2)}{x} + \tilde f^\perp(x,\bpt^2) , \\
g_T^\prime(x,\bpt^2) & = & \frac{m}{M}\,\frac{h_{1T}(x,\bpt^2)}{x}
+ \tilde g_T^\prime(x,\bpt^2) , \\
g_L^\perp(x,\bpt^2) & = & \frac{g_{1L}(x,\bpt^2)}{x} +
\frac{m}{M}\,\frac{h_{1L}^\perp(x,\bpt^2)}{x} + \tilde g_L^\perp(x,\bpt^2) , \\
g_T^\perp(x,\bpt^2) & = & \frac{g_{1T}(x,\bpt^2)}{x} +
\frac{m}{M}\,\frac{h_{1T}^\perp(x,\bpt^2)}{x} + \tilde g_T^\perp(x,\bpt^2), \\
h_T^\perp(x,\bpt^2) & = & \frac{h_{1T}(x,\bpt^2)}{x} 
+ \tilde h_T^\perp(x,\bpt^2),\\
h_L(x,\bpt^2) & = & \frac{m}{M}\,\frac{g_{1L}(x,\bpt^2)}{x}
- \frac{\bpt^2}{M^2}\,\frac{h_{1L}^\perp(x,\bpt^2)}{x} 
+ \tilde h_L(x,\bpt^2), \\
h_T(x,\bpt^2) & = & \frac{m}{M}\,\frac{g_{1T}(x,\bpt^2)}{x}
- \frac{h_{1T}(x,\bpt^2)}{x} 
- \frac{\bpt^2}{M^2}\,\frac{h_{1T}^\perp(x,\bpt^2)}{x}
+ \tilde h_T(x,\bpt^2).
\ea
The $\bpt$-integrated results involve the functions
\ba
e(x) & \equiv & \int d^2\bpt\, e(x,\bpt^2)
= \frac{m}{M}\,\frac{f_1(x)}{x} + \tilde e(x), \\
g_T(x) & \equiv & \int d^2\bpt\,\left[ g_T^\prime(x,\bpt^2)
+ \frac{\bpt^2}{2M^2}\, g_T^\perp (x,\bpt^2) \right]  
= \frac{g_{1T}^{(1)} (x)}{x} + \frac{m}{M} \frac{h_1(x)}{x} 
+ \tilde g_T(x), \\
h_L(x) & \equiv & \int d^2\bpt\,h_L(x,\bpt^2) 
= - 2\,\frac{h_{1L}^{\perp (1)} (x)}{x} + \frac{m}{M} \frac{g_1(x)}{x} 
+ \tilde h_L(x),
\ea
where the functions $\tilde e(x)$, $\tilde g_T(x)$ and $\tilde h_L(x)$
are the $\bpt$-integrated results of the interaction-dependent parts.
The functions with upper index $(1)$ indicate $\bpt^2/2M^2$-weighted functions,
\ba
g_{1T}^{(1)}(x) 
\equiv \int d^2\bpt\,\frac{\bpt^2}{2M^2}\,g_{1T}(x,\bpt^2), \\
h_{1L}^{\perp (1)}(x) 
\equiv \int d^2\bpt\,\frac{\bpt^2}{2M^2}\,h_{1L}^\perp(x,\bpt^2). 
\ea
Using the explicit expansion into amplitudes one can derive the following
relations for the $p_T$ integrated distribution functions
\cite{Tangerman-Mulders-94},
which clearly exhibit the role of transverse momenta and interaction terms,
\ba
g_T(x) & = & \int_x^1 dy \,\frac{g_1(y)}{y}
+ \frac{m}{M} \left[ \frac{h_1(x)}{x} - \int_x^1 dy\, \frac{h_1(y)}{y^2} \right]
+ \tilde g_T(x) - \int_x^1 dy\, \frac{\tilde g_T (y)}{y}, \nonumber \\
& = & g_1(x) + \frac{d}{d x} \,g_{1T}^{(1)} (x) 
\label{gti}
\\
h_L(x) & = & 2x\int_x^1 dy\,\frac{h_1(y)}{y^2}
+ \frac{m}{M} \left[ \frac{g_1(x)}{x}
- 2x \int_x^1 dy\,\frac{g_1(y)}{y^3} \right]
+ \tilde h_L(x) - 2x \int_x^1 dy\,\frac{\tilde h_L(y)}{y^2} \nonumber \\
& = & h_1(x) - 2\,\frac{d}{d x}\,h_{1L}^{\perp (1)} (x).
\ea
The decomposition of $g_T$ contains
the Wandzura-Wilczek part \cite{Wandzura-Wilczek-77}, a quark mass term
and interaction-dependent parts. Noteworthy is the appearance of the
$(\bpt^2/2M^2)$-weighted twist-two functions, e.g.\ the function
$g_{1T}^{(1)}(x)$ [In Ref.~\cite{Ratcliffe-86} referred to as $B^A(x)$].
 
For the fragmentation part one has for the projections
\be
\Delta_A^{\alpha [\Gamma]}(z,\bkt) = \Delta_D^{\alpha [\Gamma]}(z,\bkt) 
- k^\alpha\,\Delta^{[\Gamma]}(z,\bkt),
\ee
the result $\Delta_A^{- [\Gamma]} = 0$, while for transverse indices
\be
\Delta_A^{\alpha [\Gamma ]}(z,\bkt) =
\left. \int \frac{d\xi^+ d^2\xi_\perp}{4z\,(2\pi)^3} \ e^{i\,k\cdot \xi}\,
Tr\,\langle 0 \vert \psi(\xi)
\, A^\alpha (x)\,a_h^\dagger a_h\, \overline \psi(0) \Gamma \vert 0 
\rangle \right|_{x^- \,=\, 0} .
\ee
It is convenient to parametrize the following combinations (in which all
indices $\alpha$ and $\beta$ are transverse only) as
\begin{eqnarray}
&&\Delta_A^{\alpha [i\sigma^{\beta -}\gamma_5]}-\Delta_A^{\beta
[i\sigma^{\alpha -}\gamma_5]} = 
-i\epsilon_{\scriptscriptstyle T}^{\alpha \beta} \left(\frac{M_h}{z}
\htE -i\,\frac{M_h}{z}\,\htH  \right)
\nonumber \\ & & \qquad \qquad \qquad \qquad \qquad \qquad \quad
- \left( S_{hT}^\alpha k_T^\beta - k_T^\alpha S_{hT}^\beta
\right)\, (\frac{1}{z}\,\htH_T^\perp +i\,\frac{m}{M_h}
\,\htF_{1T}^\perp),
\\ & &
g_{{\scriptscriptstyle T}\alpha\beta}\,
\Delta_{A}^{\alpha [i\sigma^{\beta -}\gamma_5 ]}
=  \frac{M_h}{z} \,\htH_s + i\,\frac{M_h}{z}\,\htE_s ,
\\ & &
\Delta_A^{\alpha [\gamma^{-}]} 
+ i\epsilon_{\scriptscriptstyle T}^{\alpha \beta}\,
\Delta_{A\,\beta}^{\ [\gamma^{-}\gamma_5]}
= k_T^\alpha \left(\frac{1}{z}\,\htF^\perp 
+i\,\frac{m}{M_h}\,\htH_1^\perp\right)
-\frac{k_T^\alpha}{M_h}\,\epsilon_{{\scriptscriptstyle T}\,ij}
k_T^i S_{hT}^j\,\htF_{1T}^\perp
\nonumber \\ & & \qquad \qquad \qquad \qquad \qquad \qquad \quad
+ i\epsilon_{\scriptscriptstyle T}^{\alpha \beta}k_{T \beta}
\,\frac{1}{z}\,\htG_s^\perp 
+ i\epsilon_{\scriptscriptstyle T}^{\alpha \beta}S_{hT\beta} 
\,\frac{M_h}{z}\,\htG_T^\prime,
\end{eqnarray}
in terms of (interaction-dependent) functions indicated with a tilde
and depending on $z$ and $\bkt^\prime$ = $-z\bkt$. One
again can use them to split the fragmentation functions into a 
'parton' piece and interaction dependent functions,
\begin{eqnarray}
\hF_{1T}^\perp(z,\bkt^{\prime 2}) &=& \htF_{1T}^\perp(z,\bkt^{\prime 2}),\\
\hH_1^\perp(z,\bkt^{\prime 2}) &=& \htH_1^\perp(z,\bkt^{\prime 2}),\\
\hE(z,\bkt^{\prime 2}) &=& \frac{m}{M_h}\,z\hF_1(z,\bkt^{\prime 2}) 
+ \htE(z,\bkt^{\prime 2}),\\
\hF^\perp(z,\bkt^{\prime 2}) &=& z\hF_1(z,\bkt^{\prime 2}) 
+ \htF^\perp(z,\bkt^{\prime 2}),\\
\hE_s(z,\bkt^\prime) &=& \htE_s(z,\bkt^\prime),\\
\hG_T^\prime(z,\bkt^{\prime 2}) &=& \frac{m}{M_h}\,z\hH_{1T}(z,\bkt^{\prime 2}) 
+ \htG_T^\prime(z,\bkt^{\prime 2}),\\
\hG_s^\perp(z,\bkt^\prime) &=& z\hG_{1s}(z,\bkt^\prime) 
+ \frac{m}{M_h}\,z\hH_{1s}^\perp(z,\bkt^\prime) 
+ \htG_s^\perp(z,\bkt^\prime),\\
\hH_T^\perp(z,\bkt^{\prime 2}) &=& z\hH_{1T}(z,\bkt^{\prime 2}) 
+ \htH_T^\perp(z,\bkt^{\prime 2}),\\
\hH(z,\bkt^{\prime 2}) &=& 
-\frac{\bkt^2}{M_h^2}\,z\hH_1^\perp(z,\bkt^{\prime 2}) 
+ \htH(z,\bkt^{\prime 2}),\\
\hH_s(z,\bkt^\prime) &=& \frac{m}{M_h}\,z\hG_{1s}(z,\bkt^\prime)
-\frac{\bkt\cdot \bS_{hT}}{M_h}\,z\hH_{1T}(z,\bkt^{\prime 2})
-\frac{\bkt^2}{M_h^2}\,z\hH_{1s}^\perp(z,\bkt^\prime) + \htH_s(z,\bkt^\prime).
\end{eqnarray}
Note that the twist-two time-reversal odd functions $\hF_{1T}^\perp$ and 
$\hH_1^\perp$ are interaction-dependent.
Below we give some of the $\bkt^\prime$-integrated functions that turn out
to be relevant in the integrated cross sections. 
They are
\ba
\hE(z) & \equiv & \int d^2\bkt^\prime \, \hE(z,\bkt^{\prime 2})
= z^2\int d^2\bkt\, \hE(z,z^2\bkt^2)
= \frac{m}{M_h}\,z\hF_1(z) + \htE(z), \\
\hG_T(z) & \equiv & \int d^2\bkt^\prime\,\left[ \hG_T^\prime(z,\bkt^{\prime 2})
+ \frac{\bkt^2}{2M_h^2}\, \hG_T^\perp (z,\bkt^{\prime 2}) \right]  
= z\hG_{1T}^{(1)} (z) + \frac{m}{M_h} \,z\hH_1(z) + \htG_T(z), \\
\hH_L(z) & \equiv & \int d^2\bkt^\prime\,\hH_L(z,\bkt^{\prime 2}) 
= - 2z\hH_{1L}^{\perp (1)} (z) + \frac{m}{M_h} \,z\hG_1(z) + \htH_L(z), 
\ea
while the $\bkt^2/2M_h^2$-weighted functions (indicated with index $(1)$) are,
\ba
\hF_{1T}^{\perp (1)}(z)  & \equiv &
\int d^2\bkt^\prime 
\,\frac{\bkt^2}{2M_h^2}\,\hF_{1T}^\perp(z,\bkt^{\prime 2}), \\
\hG_{1T}^{(1)}(z)  & \equiv &
\int d^2\bkt^\prime 
\,\frac{\bkt^2}{2M_h^2}\,\hG_{1T}(z,\bkt^{\prime 2}), \\
\hH_{1L}^{\perp (1)}(z)  & \equiv &
\int d^2\bkt^\prime 
\,\frac{\bkt^2}{2M_h^2}\,\hH_{1L}^\perp(z,\bkt^{\prime 2}).
\ea
The above decomposition leads 
in combination with the amplitude expansion (based on Lorentz invariance)
to the following relations,
\ba
\hG_T(z) & = & z\int_z^1 dy \,\frac{\hG_1(y)}{y^2}
+ \frac{m}{M_h} \left[ z\hH_1(z) - z\int_z^1 dy\, \frac{\hH_1(y)}{y} \right]
+ \htG_T(z) - z\int_z^1 dy\, \frac{\htG_T (y)}{y^2}, \nonumber \\
& = & \hG_1(z) + z\,\frac{d}{d z} \left( z\hG_{1T}^{(1)} (z)
\right), \\
\hH_L(z) & = & 2\int_z^1 dy\,\frac{\hH_1(y)}{y}
+ \frac{m}{M_h} \left[ z\hG_1(z)
- 2 \int_z^1 dy\,\hG_1(y) \right]
+ \htH_L(z) - 2 \int_z^1 dy\,\frac{\htH_L(y)}{y} \nonumber \\
& = & \hH_1(z) - 2z\,\frac{d}{d z} \left(
z\hH_{1L}^{\perp (1)} (z)\right).
\ea
Analogous to the sum rules $\int_0^1 dx\,g_2(x)$ = 0 
\cite{Burkhardt-Cottingham-76}, where $g_2 = g_T - g_1$,
and $\int_0^1 dx\,h_2(x)$ = 0 \cite{Burkardt-93,Tangerman-Mulders-94}, 
where $h_2 = 2(h_L - h_1)$, one obtains sum rules
\be
\int_0^1 dz\,\frac{\hG_2(z)}{z} = 0, 
\ee
where $\hG_2 = \hG_T - \hG_1$ and
\be
\int_0^1 dz\,\frac{\hH_2(z)}{z} = 0, 
\ee
where $\hH_2 = 2(\hH_L - \hH_1)$. As for the distribution functions, one must be
aware of convergence problems in the integrals and of the fact that in
the cross sections other contributions beyond tree-level may be present.  
We consider these sum rules at this point as merely academic.

\section{Convolutions for gaussian distributions}
 
In order to study the behavior of the convolutions of distribution and
fragmentation functions it is useful to consider gaussian distributions,
\ba
f(x,\bpt^2) & = & f(x,0)\,\exp (-R_H^2 \bpt^2)
\nonumber \\ & = & f(x)\,\frac{R_H^2}{\pi}\,\exp (-R_H^2 \bpt^2)
\equiv f(x)\,{\cal G}(\vert\bpt\vert;R_H), \\
\hF(z,\bkt^{\prime 2}) & = & \hF(z,0)\,\exp (-R_h^2 \bkt^2).
\nonumber \\ & = & \hF(z)\,\frac{R_h^2}{\pi\,z^2}\,\exp (-R_h^2 \bkt^2)
= \frac{\hF(z)}{z^2}\,{\cal G}(\vert\bkt\vert;R_h) 
= \hF(z)\,{\cal G}\left(z\vert\bkt\vert ; \frac{R_h}{z}\right).
\ea
In that case the convolution becomes
\ba
I[f\,\hF](x,z,Q_T) & = & \int d^2\bpt
\,f(x,\bpt^2)\,\hF(z,\vert \bP_{h\perp}-z\bpt\vert^2) \nonumber \\
& = & \frac{\pi}{R_H^2 + R_h^2}\,\exp
\left\lgroup - \frac{Q_T^2\,R_H^2 R_h^2}{R_H^2 + R_h^2} \right\rgroup
f(x,\bnull_T)\,\hF(z,\bnull_T)\nonumber \\
& = & f(x)\,\hF(z)\,\frac{{\cal G}(Q_T;R)}{z^2},
\ea
where $R^2$ = $R_H^2\,R_h^2/(R_H^2 + R_h^2)$.
The other convolutions that appear in the cross sections are of the form
{
\ba
&& I\Bigl[ \frac{\bhh\cdot \bpt}{M}\,f\,\hF \Bigr]
\ = \ \frac{Q_T\,R^2}{M\,R_H^2}\,I[f\,\hF], \\
&& I\Bigl[ \frac{\bhh\cdot \bkt}{M_h}\,f\,\hF \Bigr]
\ = \ -\frac{Q_T\,R^2}{M_h\,R_h^2}\,I[f\,\hF], \\
&& I\Bigl[ \frac{\bpt\cdot \bkt}{MM_h}\,f\,\hF \Bigr]
\ = \ \frac{R^2}{MM_h\,R_H^2\,R_h^2}
\left( 1 - Q_T^2\, R^2 \right)\,I[f\,\hF], \\
&& I\Bigl[ \frac{\bhh\cdot \bkt}{M_h}\,\frac{\bpt^2}{M^2}\,f\,\hF\Bigr]
\ = \ \frac{Q_T\,R^6}{M_h\,M^2\,R_h^4\,R_H^4}\,\left(
\frac{R_h^2-R_H^2}{R^2} - Q_T^2\,R_h^2 \right)\,I[f\,\hF], \\
&& I\Bigl[ \frac{2(\bhh\cdot\bpt)^2 - \bpt^2}{M^2}\,f\,\hF \Bigr]
\ = \ \frac{Q_T^2\,R^4}{M^2\,R_H^4}\,I[f\,\hF], \\
&& I\Bigl[ \frac{2(\bhh\cdot\bkt)^2 - \bkt^2}{M_h^2}\,f\,\hF \Bigr]
\ = \ \frac{Q_T^2\,R^4}{M_h^2\,R_h^4}\,I[f\,\hF], \\
&& I\Bigl[ \frac{2(\bhh\cdot\bpt)\,(\bhh\cdot \bkt) -
\bpt\cdot\bkt}{MM_h}\,f\,\hF \Bigr]
\ = \  -\frac{Q_T^2\,R^4}{MM_h\,R_H^2\,R_h^2}\,I[f\,\hF], \\
&& I\Bigl[ \frac{2(\bhh\cdot\bpt)\,(\bhh\cdot \bkt) -
\bpt\cdot\bkt}{MM_h}\,\frac{\bpt^2}{M^2}\,f\,\hF \Bigr]
\ = \  -\frac{Q_T^2\,R^8}{M^3M_h\,R_H^6\,R_h^4}\,
\left(\frac{2R_H^2 - R_h^2}{R^2} + Q_T^2\,R_h^2 \right)\,I[f\,\hF], \\
&& I\Bigl[ \frac{4(\bhh\cdot\bpt)^2(\bhh\cdot \bkt)
- 2(\bhh\cdot\bpt)\,(\bpt\cdot \bkt)
- \bp_\perp^2\,(\bhh \cdot \bkt)}{M^2M_h}\,f\,\hF \Bigr]
\ = \ -\frac{Q_T^3\,R^6}{M^2M_h\,R_H^4\,R_h^2}\,I[f\,\hF].
\ea }
Note that we have suppressed possible dependence of the radii on the specific
function or the kinematic variables irrelevant in the integration over 
transverse momenta. We can trivially generalize the results by 
taking into account such dependence, $R_H \rightarrow R_H^{f}(x)$ and
$R_h \rightarrow R_h^{\hF}(z)$.

\newpage
 
\begin{table}[b]
\caption{\label{contractions}
Contractions of the lepton tensor $L_{\mu \nu}$ with tensor
structures appearing in the hadron tensor.}
\begin{tabular}{cl}
$w^{\mu \nu}$ & $L_{\mu \nu} w^{\mu \nu}$ \\ 
& \\ \hline
$-g_\perp^{\mu \nu}$ &
$\frac{4 Q^2}{y^2} \left( 1 - y + \frac{1}{2}y^2 \right)$ \\
& \\
$a_\perp^{\,\{ \mu} b_\perp^{\nu\}} -
(a_\perp \cdot b_\perp)\,g_\perp^{\mu\nu}$ &
$\frac{4 Q^2}{y^2} \left( 1 - y \right)
\vert \baa_\perp \vert \, \vert \bbb_\perp \vert\,\cos(\phi_a+\phi_b)$ \\
& \\
$\frac{1}{2} \left(
a_\perp^{\{\mu}\,\epsilon_\perp^{\nu \} \rho} b_{\perp\rho}
+ b_\perp^{\{\mu}\,\epsilon_\perp^{\nu \} \rho} a_{\perp\rho}\right)$ &
$-\frac{4 Q^2}{y^2} \left( 1 - y \right)
\vert \baa_\perp \vert \, \vert \bbb_\perp \vert\,\sin(\phi_a+\phi_b)$ \\
= $a_\perp^{\{\mu}\,\epsilon_\perp^{\nu\} \rho} b_{\perp\rho}
-(\epsilon_\perp^{\rho\sigma}a_{\perp \rho} b_{\perp\sigma})\,g_\perp^{\mu\nu}$
& \\ & \\
$\hat t^{\,\{\mu}a_\perp^{\nu\}}$ &
$-\frac{4 Q^2}{y^2}( 2 - y)\sqrt{1-y}\,\vert \baa_\perp \vert\,\cos \phi_a$ \\
& \\
$\hat t^{\,\{\mu }\,\epsilon_\perp^{\nu\} \rho}\,a_{\perp\rho}$ &
$\frac{4 Q^2}{y^2}( 2 - y)\sqrt{1-y}\,\vert \baa_\perp \vert\,\sin \phi_a$ \\
& \\
$i\,\epsilon_\perp^{\mu \nu}$ &
$\lambda_e\,\frac{4Q^2}{y^2}\,y\left(1-\frac{y}{2}\right)$ \\
& \\
$i\,a_\perp^{\,[ \mu} b_\perp^{\nu ]}$ &
$\lambda_e\,\frac{4Q^2}{y^2}\,y\left(1-\frac{y}{2}\right)\,
\vert \baa_\perp \vert \, \vert \bbb_\perp \vert\,\sin (\phi_b - \phi_a)$ \\
& \\
$i\,\hat t^{\,[ \mu} a_\perp^{\nu ]}$ &
$-\lambda_e\,\frac{4Q^2}{y^2}\,y\sqrt{1-y}\,
\vert \baa_\perp \vert \,\sin \phi_a$ \\
& \\
$i\,\hat t^{\,[\mu}\,\epsilon_\perp^{\nu\,] \rho}a_{\perp\rho}$ &
$-\lambda_e\,\frac{4Q^2}{y^2}\,y\sqrt{1-y}\,
\vert \baa_\perp \vert \,\cos \phi_a$ \\
\end{tabular}
\end{table}

\newpage

\begin{figure}[htb]
\label{fig1}
\begin{center}
\leavevmode
\epsfxsize=10 cm
\epsfbox{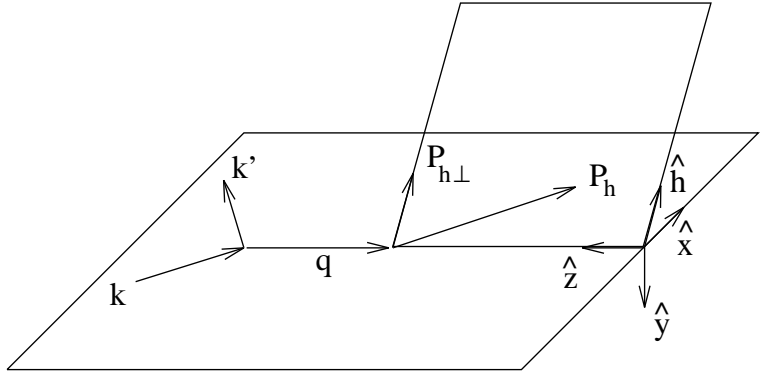}
\end{center}
\caption{Kinematics for $\ell H$ scattering.}
\end{figure}

\begin{figure}[hbt]
\label{fig2}
\begin{center}
\leavevmode
\epsfxsize=14 cm
\epsfbox{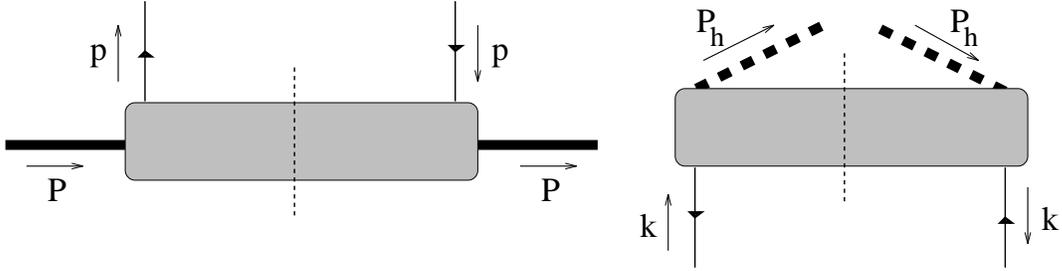}
\end{center}
\caption{Quark-quark correlation functions $\Phi(P,S;p)$ (left) and
$\Delta(P_h,S_h;k)$ (right) giving quark distributions
and fragmentation functions, respectively.}
\end{figure}

\begin{figure}[hbt]
\label{fig3}
\begin{center}
\leavevmode
\epsfxsize=14 cm
\epsfbox{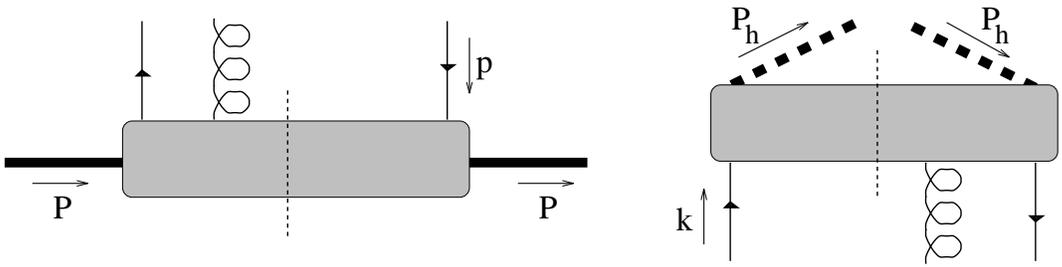}
\end{center}
\caption{Quark-quark-gluon correlation functions $\Phi_A$ (left)
and $\Delta_A$ (right) contributing in hard
scattering processes at subleading order.}
\end{figure}

\begin{figure}[htb]
\label{fig4}
\begin{center}
\leavevmode
\epsfxsize=7 cm
\epsfbox{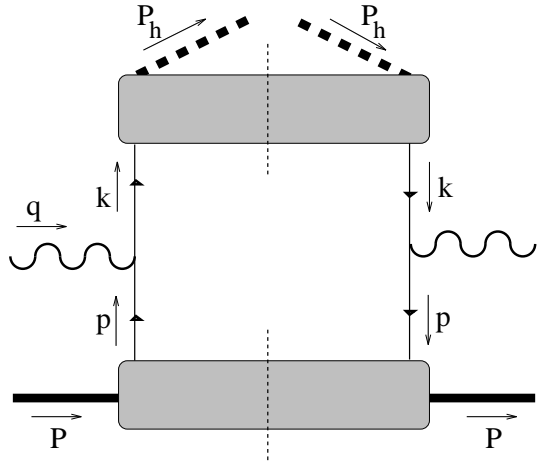}
\end{center}
\caption{Quark diagram contributing to $\ell H$ scattering in leading order.
There is a similar antiquark contribution.}
\end{figure}

\begin{figure}[htb]
\label{fig5}
\begin{center}
\leavevmode
\epsfxsize=16 cm
\epsfbox{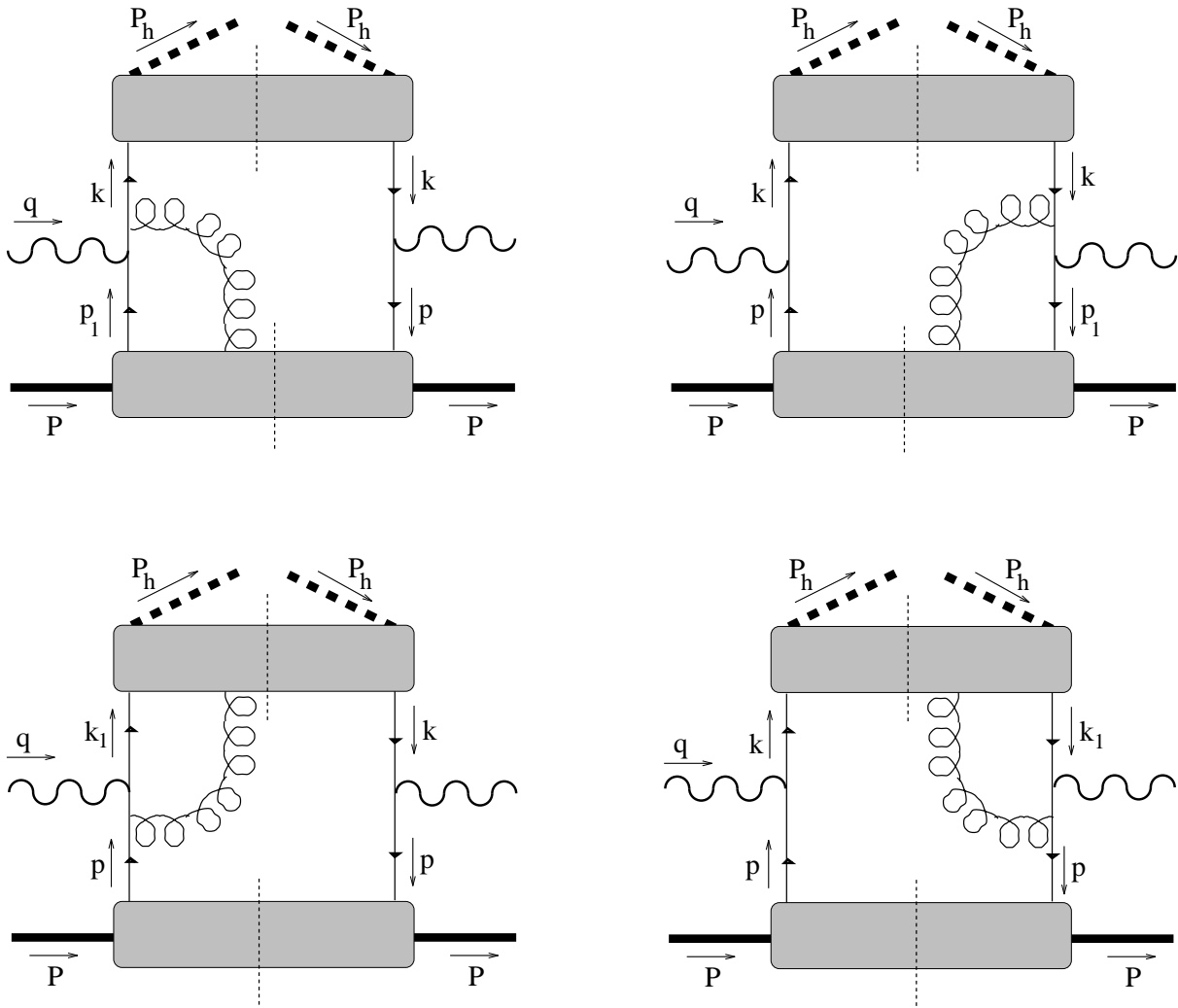}
\end{center}
\caption{Quark diagrams contributing to $\ell H$ scattering at order $1/Q$.}
\end{figure}

 
\end{document}